\documentclass[prb,twocolumn,10pt,aps,longbibliography,superscriptaddress]{revtex4-2}

\usepackage[english]{babel}
\usepackage{bm} 
\usepackage{amsmath}
\usepackage{amssymb}
\usepackage{amsfonts}
\usepackage{mathrsfs}
\usepackage{comment}
\usepackage{graphicx}
\usepackage{color}
\usepackage{dcolumn} 
\usepackage{soul} 
\usepackage{multirow}
\usepackage{cancel}
\usepackage{comment} 
\usepackage{epstopdf} 

\begin{document}
\title{
Smoothed-Cubic Spin-Glass Model of Random Lasers
}

\author{Marcello Benedetti$^{1,2}$ and Luca Leuzzi}
\email{luca.leuzzi@cnr.it}
\affiliation{Dipartimento di Fisica, Universit\`a di Roma
  ``Sapienza'', Piazzale A. Moro 5, I-00185, Roma, Italy}
\affiliation{Institute of Nanotechnology of the National Research Council of Italy, CNR-NANOTEC Roma, Piazzale A. Moro 5, I-00185, Rome, Italy}

\begin{abstract}
We study the equilibrium  glassy behavior of a multimode random laser model with nonlinear four-body quenched disordered interactions and a global smoothed-cubic constraint on mode intensities. 
This constraint, which provides a more realistic representation of gain saturation than the commonly used spherical constraint, prevents intensity condensation while preserving the dense, long-range interaction structure characteristic of many multistate random lasers.
The model effective  Hamiltonian  is a function of mode amplitudes with random frequencies and is defined on a complete mode-locked graph.  Using large-scale GPU-accelerated Monte Carlo simulations with the Parallel Tempering algorithm, we analyze systems of varying sizes to probe their thermodynamic-limit behavior. Finite-size scaling of the specific heat, of the Parisi overlap distributions, and of the inverse participation ratio's reveals a spin-glass transition, with critical exponents matching the mean-field Random Energy Model universality class. The smoothed-cubic constraint produces broad, non-condensed intensity distributions, avoiding the pseudo-condensation seen in spherical models on the same interaction   graph. Our results show that more realistic gain-saturation constraints preserve spin-glass characteristics while enabling simulations of larger, more dilute systems, providing a robust framework for studying glassy random lasers with self-starting mode-locking.
\end{abstract}

\maketitle

\section{Introduction}

If enough energy is pumped into a disordered, optically active medium, multiple scattering can induce population inversion in atoms or molecules above an optical gap, producing a random laser \cite{Cao98,Cao99,Cao00,Anni04,Wiersma08,vanderMolen07,Tulek10,Andreasen11,Folli12,Folli13,Viola18,Antenucci21,Gomes21,Elizer22,Xia2023,Du2024,Camara2024,Yadav2025,Qi2025,Qiang2025,Vitiello2025,Yadav2025,Zhang2025}.
Very schematically we can say that random lasers consist of two primary elements:
(i) an optically active (gain) medium, which provides amplification, and (ii)
randomly placed scatterers, which offer a high refractive index and the feedback needed for stimulated emission.
Depending on the specific properties of the random lasing compound, the two elements can be combined into a single component.
These systems can be built in very different ways, can be both solid or liquid, 2D or 3D, organic or inorganic, the optically active material can be confined or spread all over the volume. 
Random lasers are  multimode, that is, they emit on different resonances. 
Some  may show multiple sub-nanometer spectral peaks 
above a pump threshold~\cite{Cao99}, whereas others display smoother, though always disordered, emission spectra. 

Unlike
standard multimode lasers employed to yield ultrafast pulses, random lasers do not require precise alignment and complex designs. Indeed, though less powerful and precise, they are simpler to construct, cost-effective, and highly flexible. Their emission is usually non-directional, and they enable diverse applications, including
speckle-free imaging \cite{Redding12,Barredo-Zuriarrain17},
    granular matter studies \cite{Folli12,Folli13},
    remote sensing \cite{Ignesti16,Xu17,Gomes21},
 medical diagnostics and  imaging \cite{Polson04,Song10,Lahoz15,Wang17,Gomes21}, or
    optical amplification and optoelectronic devices \cite{Lin12,Liao16,Gomes21}, just to mention a few.
    

Depending on the material, particularly on its optical and scattering properties, random spectral fluctuations between different pumping shots (i.~e., different realizations of the same random laser, else called {\it real replicas}) may or may not vary significantly.
A wide variety of  spectral features is reported, e. g., in Refs. 
\cite{Turitsyn10,Leonetti11,Folli13,Baudouin13,Antenucci21}, depending on material compounds and experimental setups. 
Random lasers are open systems, that is,  rather than oscillating between well specific boundaries (mirrors) as in standard lasers, light can propagate and be emitted in any direction. Therefore, emission acquisition can be a delicate issue, as well.
 Finally, also the scattering strength and the pumping conditions   affect the emission.

 
In the last years experiments on a certain class of random lasers  provided evidence of particularly
non-trivial correlations between the  shot-to-shot
fluctuations of the emission
spectra. These are now termed {\it glassy} random lasers~\cite{Ghofraniha15,Antenucci16,Gomes16,Pincheira16, Basak16,Lopez18,Gomes21,Qiang2025,Qi2025} because of an analogy to multi-equilibria physics in glassy systems \cite{Angelani06a,Leuzzi09a,Antenucci15a}. 
These special correlations are predicted by a  theory based on statistical mechanics of  complex disordered systems ~\cite{Antenucci15a,Antenucci15b,Antenucci16,Antenucci16b}. Indeed, it  has been shown that 
these fluctuations are compatible with an
organization of mode configurations in clusters of states, similar to
the one occurring in complex disordered systems such as spin glasses.  Such a
correspondence has been analytically  proved in a spherical mixed multi-spin model in the narrow-band approximation \cite{Siegman1986}, proving the
equivalence between the distribution of the 
overlap between  intensity fluctuations in the emission spectra and the distribution of the overlap between states, the    
so-called Parisi overlap, the order parameter of the glass
transition\cite{Antenucci15f}.

 In  models for random multimode lasers with distinct frequencies \cite{Gradenigo20,Niedda23a,Niedda23b}  the four-waves non-linear mixing between
electromagnetic field modes is controlled by a constraint on the relative mode frequencies, termed {\it mode-locking} (ML). 
 In ML  lasers, effective mode-interactions occur only for quadruplets of modes with constructive interference; that is, modes 
whose frequencies $\omega_k$ satisfy
the condition
\begin{equation} 
|\omega_{k_1}-\omega_{k_2}+\omega_{k_3}-\omega_{k_4}| < \gamma,
\label{eq:selection-rule}
\end{equation}
with $\gamma$ being the typical line-width of the modes. We will refer to Eq.~(\ref{eq:selection-rule}) as Frequency Matching
Condition (FMC). In standard ML  lasers such selection rule is implemented by ad hoc nonlinear devices (e.g., saturable absorbers for passive mode-locking~\cite{Haus00}) that are not there in random lasers. 
As hypothesized in Ref.~\cite{Conti11}, and  experimentally demonstrated in Ref.~\cite{Antenucci21}, though,  in these systems mode-locking  occurs as a self-starting phenomenon. 
 We  call an
interaction network built on the ML  selection rule in
Eq.~(\ref{eq:selection-rule}) a {\it mode-locking graph}.

From the point of view of statistical mechanics of complex disordered systems, random lasers represent, so far, the only physical system where
  the relevant degrees of freedom, namely the complex amplitudes of the light
  modes, naturally
  form a dense long-range interaction network of the kind for which replica symmetry breaking  mean-field
  theory \cite{Mezard87} is proved to work exactly, as in high dimension spin-glasses \cite{Parisi80,Parisi83} or structural glasses made of hard spheres \cite{PUZ}. It is not by chance that random lasers have been the first, and to our knowledge so far the only,
 complex disordered system with a large number of degrees of freedom providing experimental
  evidence of a  replica symmetry-breaking
pattern~\cite{Ghofraniha15,Gomes16,Pincheira16, Basak16,Lopez18,Gomes21,Chen2025}.

Though the system is mean-field, introducing a non-trivial mode-locking and mode constraints more realistic than spherical, does not allow for an analytic approach. Therefore, in this work we resort to (highly optimzed) Monte Carlo simulations of the dynamics.
The outline of this work is the following. 
In Sec. II we recall the effective theory of the random laser dynamics and we introduce the amplitude model with a global {\it smoothed cubic} constraint, rather than spherical. We also introduce a random distribution for the mode frequencies and adapt the FMC, see Eq. (\ref{eq:gamma_scaling}), to perform meaningful finite size scaling analysis from numerical data.
In Sec. III we describe the Monte Carlo algorithm that we have been implementing in order to simulate continuous spin, four-body, fully connected, bond-disordered systems.
In Sec. IV we report all the observables that we measure, our analysis and our results. 
In Sec. \ref{sec:conc} we provide a final discussion and conclusion and we outline possible future developments.

 \section{Modeling random laser dynamics}

Based on the principles of quantum electrodynamics in an open system, the classical stochastic dynamics of random lasers is derived from the underlying quantum many-body interactions of light and matter.
The core of the theory relies on a key simplification: in  laser media, the timescales for atomic processes  are significantly faster than the lifetimes of the standing light modes within the resonator. This separation of timescales allows for the adiabatic elimination of the atomic operators (matter fields). Further on, the  quantum dynamics can be reduced to a more manageable system of classical, non-linear stochastic equations that describe only the electromagnetic field.

 This effective theory provides the foundation for applying  statistical physics models to understand the behavior of random lasers.
A full
account of the derivation 
 from the quantum many-body
dynamics of light coupled with matter can be found in Ref.~\cite{Antenucci16}, as well as in Refs.~\cite{Antenucci15e,Niedda23a,Niedda23b}.
Here we only mention that in the classic effective representation the dynamic variables are the complex numbers $a_k(t) = A_k(t)e^{\imath\phi_k(t)}$, i.e.,  the  amplitudes of the superposition of the light normal modes comprising the discrete spectrum of the electromagnetic field inside the material:
\begin{align} \label{EMfield_normal_modes}
\bm{E}(\bm{r},t) = \sum_{k=1}^N a_k(t)e^{\imath \omega_k t}\bm{E}_k(\bm{r}) + \text{c.c.} ,
\end{align}
where $\bm{E}_k(\bm{r})$ is the space-dependent wavefunction of the mode with frequency $\omega_k$. 
These amplitudes $a_k(t)$ are the semiclassical approximation counterpart of the original electromagnetic field creation and annihilation operators, reduced from operators to complex numbers. 

A \textit{slow} amplitude mode is defined by its dynamics occurring on a time scale longer than the mode  period, $\omega_k^{-1}$. According to the slow amplitude approximation  the rapid phases $e^{\imath \omega_k t}$ are averaged out. 
In Fourier  transform this yields $a_k(t) \simeq a_k(t,\omega)\delta(\omega - \omega_k)$. Such approximation is quite robust in the lasing regime since lasing modes are inherently slow amplitude modes, as they possess a very narrow linewidth $\gamma$ around their  frequency $\omega_k$. 

The short time averaging over the $e^{\imath \omega_k t}$ oscillations results in the FMC, Eq. (\ref{eq:selection-rule}).
 The general expression for $2n$-mode effective interaction is
\begin{align} \label{FMC}
\text{FMC}(\bm{k}): | \omega_{k_1} - \omega_{k_2} + \cdots + \omega_{k_{2n-1}} - \omega_{k_{2n}} | \lesssim \gamma ,
\end{align}
of which Eq. (\ref{eq:selection-rule}) is the case $n=2$. From the point of view of the emerging statistical physics theory of effectively interacting modes,
 the FMC acts as a selection rule on the modes participating in the interactions.

The stochastic differential equation for the time evolution of the
modes $a_k$ reads, eventually, as
\begin{align} \label{LangevinEq}
\frac{d a_{k_1}}{dt} &= \frac{\partial \mathcal H[\bm a]}{\partial \bar a_k}+ \eta_{k_1}(t),
\end{align}
where 
\begin{align} \label{Hamilt2}
\mathcal{H}[\bm{a}] &= \mathcal{H}_2[\bm{a}] + \mathcal{H}_4[\bm{a}],
\end{align}
and
\begin{eqnarray}
  \mathcal{H}_2[\bm{a}] &=& - \sum_{\bm{k} | \text{FMC}(\bm{k})} J^{(2)}_{k_1 k_2} \overline{a}_{k_1}a_{k_2} + \mbox{c.c.} \nonumber \\ 
  \mathcal{H}_4[\bm{a}] &=& - \sum_{\bm{k} |\text{FMC}(\bm{k})} J^{(4)}_{k_1 k_2 k_3 k_4} \overline{a}_{k_1}a_{k_2}\overline{a}_{k_3}a_{k_4} + \mbox{c.c.}
 \nonumber 
 \label{eq:H4}.
\end{eqnarray}

The expression $\sum_{\bm{k} | \text{FMC}(\bm{k})}$ denotes the sum over the indices $\bm k=\{k_1,\ldots,k_{2n}\}$ satisfying a FMC, Eq. (\ref{FMC}).

The noise is modeled as white noise with 
$$\langle \eta_k(t)\rangle=0 \qquad, \qquad \langle \eta_j(t)\eta_k(t')\rangle = 2T \delta_{jk} \delta(t-t').$$
In principle, the noise is actually correlated, i.e., $\langle \eta_{k_1}\eta_{k_2} \rangle \neq \delta_{k_1 k_2}$. However, it can be diagonalized by changing the basis of the dynamic variables: the decomposition of resonator modes into a slow amplitude basis is not unique \cite{Feshbach58}, and one can use this freedom to construct a basis in which the noise is uncorrelated. The price for this is the introduction of effective non-diagonal linear interactions. This is not a significant complication in the random laser case, as linear couplings already possess off-diagonal contributions that account for the openness of the cavity.
They yield different contributions possibly depending on cavity gain and losses  and atom-field interaction inside the disordered medium. The latter expression is the most relevant one in the dynamics:
\begin{align}\nonumber 
J_{k_1 k_2}^{(2)} \propto & \sum_{\alpha\beta}^{\{x,y,z\}} \int_V d \bm{r}\,  \epsilon_{\alpha \beta}(\bm{r})\ E^\alpha_{k_1}(\bm{r})\ E^\beta_{k_2}(\bm{r}),
\end{align}
where $\bm{\epsilon}(\bm{r})$ is the dielectric permittivity tensor and the integral is extended over the entire volume $V$ of the medium. 
In particular, the diagonal elements of $J_{k_1 k_2}^{(2)}$ represent the net gain curve of the medium (i.e.,~the gain reduced by the losses), which plays an important role mainly below the lasing threshold. 
We do not go into details on the representation of the linear coupling  since, in this work, we will set $\mathcal H_2=0$ in Eq. (\ref{Hamilt2}) and  we will focus on the nonlinear contribution.
Indeed, 
the term responsible for the onset of lasing as the external pumping increases  is the nonlinear $4$-mode interaction. In the rest of the present work we will  neglect the linear contribution, that amounts to neglect the effects of losses and radiation of modes outside the lasing material.

The non-linear couplings $J_{k_1 k_2 k_3 k_4}^{(4)}$ are given by the spatial overlap of the electromagnetic mode wavefunctions modulated by a non-linear optical susceptibility $\chi^{(3)}$ 
\begin{align} \label{NonLinearCoup}
J^{(4)}_{k_1 k_2 k_3 k_4} \propto &\sum_{\alpha\beta\gamma\delta}^{\{x,y,z\}}\int_V d \bm{r}\  \chi^{(3)}_{\alpha \beta \gamma \delta} (\{\omega_{\bm{k}}\}; \bm{r}) \nonumber \\
&\times E^\alpha_{k_1}(\bm{r})\ E^\beta_{k_2}(\bm{r})\ E^\gamma_{k_3}(\bm{r})\ E^\delta_{k_4}(\bm{r}),
\end{align}
where, again, the integral is over the whole volume of the medium, $E_{k}^{\alpha}(\bm r)$ is the spatial profile on the $\alpha$-coordinate of the $k$-th normal mode of the electromagnetic field and $\chi^{(3)}$ is the non-linear susceptibility tensor, depending  both on space $\bm r$ and on the frequencies of the modes overlapping in space at coordinate $\bm r$.  
In general, the  couplings are complex numbers but
in the dissipative limit \cite{Antenucci16,Niedda23a} we will consider the effective coupling coefficients $J$'s as real parameters, without loss of generality.

\subsection{A constraint to implement gain saturation}
\label{sec:gainsaturation}

The laser dynamics is brought to stationarity by gain saturation:  as the injected power is kept constant, the  atoms emitting photons decay to lower states, saturating the gain of the laser. This is why the light modes stochastic dynamics can display a stationary potential solution for its probability distribution and, therefore,  we can describe the dynamics by means of the effective Hamiltonian \eqref{Hamilt2}. Indeed, when boundaries on the total energy distributed among the system modes are added, Eq.~\eqref{LangevinEq}
eventually reaches a stationary regime.

 In fact, lasers are strongly out of equilibrium: energy is constantly pumped into the system in order to keep population inversion and stimulated emission, and in the case of cavity-less systems, such as random lasers, also compensate the continuous leakages.
However, the process of gain saturation leads to a steady state since as the laser intensity increases it depletes the population inversion faster. The rate of stimulated emission eventually matches the rate at which the pump replenishes excited atoms. At this balance point, the average intensity spread among the lasing modes stops increasing.

In order to mathematically model the gain saturation, the implementation of an overall spherical constraint  fixing the total optical intensity in the system
was first proposed for standard multimode lasers in the narrow band approximation in Refs~\cite{Gordon02,Gordon03a}, and later on implemented in Monte Carlo numerical simulations of non-trivially mode-locking ordered lasers~\cite{Antenucci15c,Antenucci15d}.

 Fixing the energy shared by the modes as $\mathcal E= \sum_{i=1}^N |a_i|^2=\epsilon N$ was also adopted to realize gain saturation and stationarity in statistical mechanical models of random lasers. This  approximation for the gain saturation allows to perform analytic computation in the narrow-band limit~\cite{Antenucci15a,Antenucci15b,Antenucci15e,Antenucci15f,Antenucci16} and has been implemented also in the numerical simulation of more realistic instances of these models, including wide spectra and non-trivial mode-locking~\cite{Gradenigo20,Niedda23a}.
Besides being drastic, however, that choice has the drawback to be feasible only for spin-glass models on $p$-body interaction graphs that are {\it fully connected}  or {\it ML  fully connected}. In the first case the total connectivity, i.e., the total number of non-zero $p$-mode couplings, grows with the number $N$ of modes like $c\sim N^{p-1}$; in the latter like $c\sim N^{p-2}$. Indeed, only in those cases the system does not condense in intensity at high pumping \cite{Niedda23b}, allowing analytic computations and Monte Carlo numerical simulations to be carried out also in the random laser regime.

The spherical constraint's requirement of strictly fixed total intensity is rather restrictive. 
In this work we will implement a different global constraint for the mode intensities that models gain saturation without directly fixing the total intensity, thereby allowing for more realistic intensity fluctuations while still preventing divergence.
This reads
\begin{align} \label{SmoothedCubic}
\sum_{k=1}^N |a_k|^\rho  = \epsilon^{\rho/2} N\quad , \quad \rho=4.
\end{align}
At difference with the constraint with $\rho=2$ of fixed total intensity, in which the modes lie on a $N$-dimensional hyper-sphere, here the mode intensities are on the surface of a  smoothed hyper-cube and the overall intensity fluctuates (though it never diverges, i.e., it saturates). 

\subsection{Equilibrium-like distribution and effective photonic temperature}

The stationary regime of the system can be described as if the system is at equilibrium with an effective thermal bath \cite{Antenucci16}, whose effective  temperature (a ``photonic'' temperature) accounts both  for the spontaneous emission rate and the gain saturation constraint. 
To see it, let us first consider the smoothed cubic constraint (\ref{SmoothedCubic}) and rescale $$ \frac{a_k}{\epsilon^{1/2}} \to a_k.$$ Eq. (\ref{SmoothedCubic}) becomes $$\sum_{k}|a_{k}|^{\rho}=N. $$ 
The effective distribution for the phasor configuration $\bm a=\{a_1,\ldots,a_N\}$ will, then, be 
\begin{eqnarray} \label{ProbDistr}
P[\bm a]&=& \frac{e^{-\beta_{\rm ph}\mathcal H[\bm a] }\delta\left(N - \sum_{k=1}^N |a_k|^4 \right)}{Z} 
\end{eqnarray}
where $\beta_{\rm ph}$ is the effective {\it photonic} inverse  temperature of the statistical mechanical model representing the stationary regime of the random laser. It depends on both (i) the heat bath temperature $T$, linked to the spontaneous emission rate through the kinetic energy of the atoms and (ii)
the energy per mode $\epsilon$ of the pumping laser. 

Eventually, the following relation holds
\begin{equation}
\label{eq:beta_ph}
    \beta_{\rm ph}\equiv \frac{\epsilon^{2}}{T }.  
\end{equation}

\begin{figure}[t!]
\includegraphics[width = .99\columnwidth]{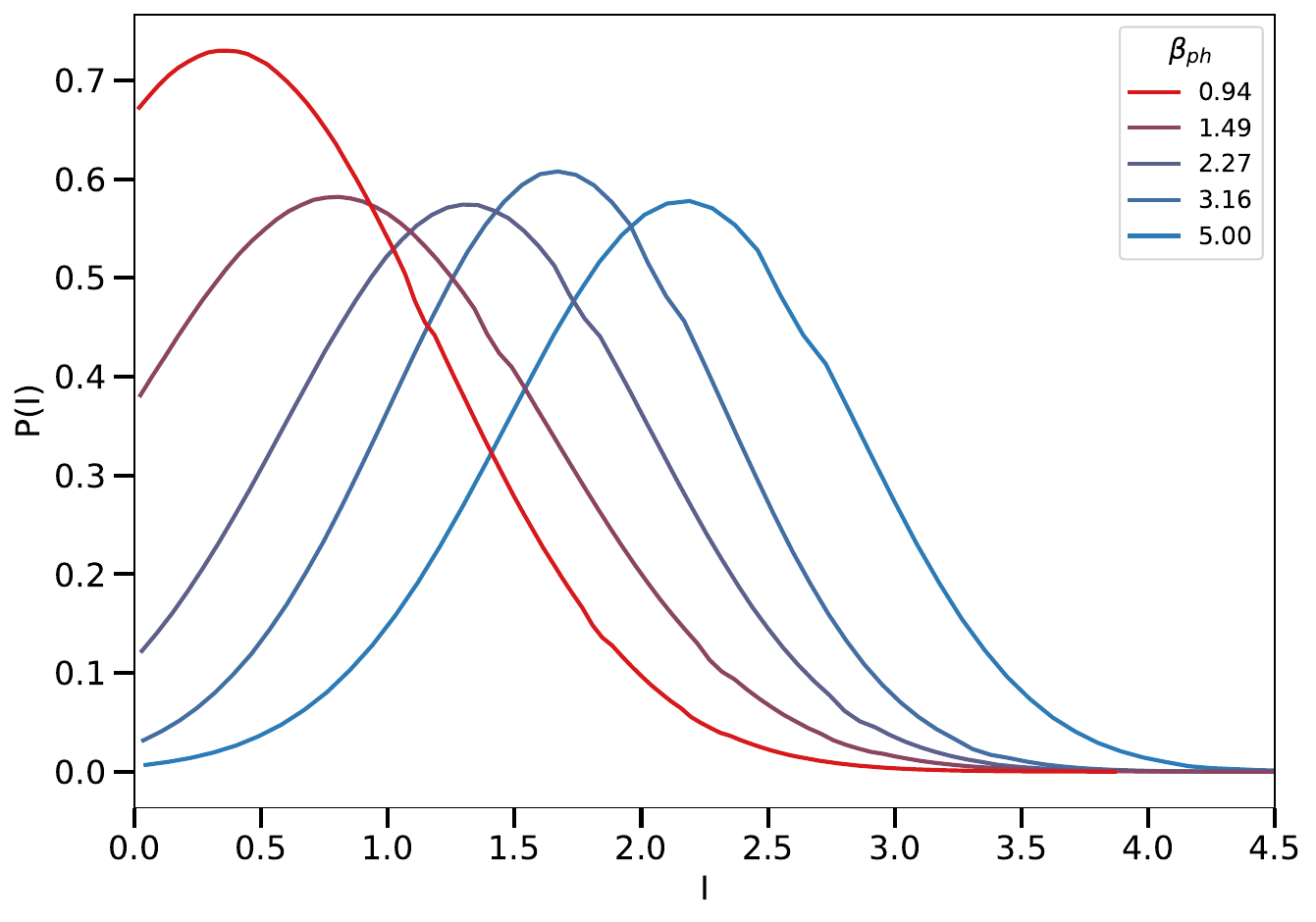}
\caption{Distribution of the system intensity in a $p=4$ multi-mode-coupling model with a smoothed-cubic global constraint \eqref{SmoothedCubic} at different (photonic) temperatures.}
\label{Fig:P_intensity}
\end{figure}

Under the smoothed-cubic constraint the overall intensity $I=\sum_{k=1}^N |a_k|^2$ is not a constant anymore but varies in time in a given range, better approximating what  occurs in real materials with gain saturation. 
In Fig. \ref{Fig:P_intensity} we display the probability distribution $P(I)$ of the global intensity at equilibrium at different photonic temperatures for a system with $N=96$ modes. 

\subsection{Mode-locking and random  mode-coupling network}
 Looking at Eq. (\ref{NonLinearCoup}), we observe that
the precise value of the couplings in the Hamiltonian Eq.~\eqref{Hamilt2} requires the knowledge of the spatial profile of the wave functions of the modes, which is not available in random lasers, since they are characterized by a complicated spatial structure of the modes, quite difficult to acquire experimentally, above all when  many competing standing modes are established in the system. We now try to give a representation of the coupling network and the coupling constant random distribution encoding the main constituent features of a random laser.

\subsubsection{Quenched random frequencies}
In order to assign a frequency $\omega_k$ to the standing mode $k$, in the numerical simulations a uniform random number $r_k$  between $0$ and $1$ is extracted.
The mode indices are assigned to the sorted list of frequencies, so that $r_1<r_2<\ldots <r_N$.
If we want 
to enforce in the   laser model with random frequencies   the same bond dilution occurring in lasers with comb-like frequency distributions~\cite{Schawlow58,Savchenkov58,Diddams00, Antenucci15e,Antenucci15c} we apply the FMC with a size-dependent linewidth 
\begin{equation}
    \label{eq:gamma_scaling}
\gamma = \frac{1}{2(N-1)}
\end{equation}
in which the $1/(N-1)$ factor takes into account that numerically we keep the frequency distribution domain in $[0,1]$, while increasing the number of modes. The factor $1/2$ is there just to  follow exactly the theoretical behavior with $N$ for comb-like systems
\cite{TrincaCintioli}.

\begin{figure}[t!]
    \centering
\includegraphics[width=0.99\linewidth]{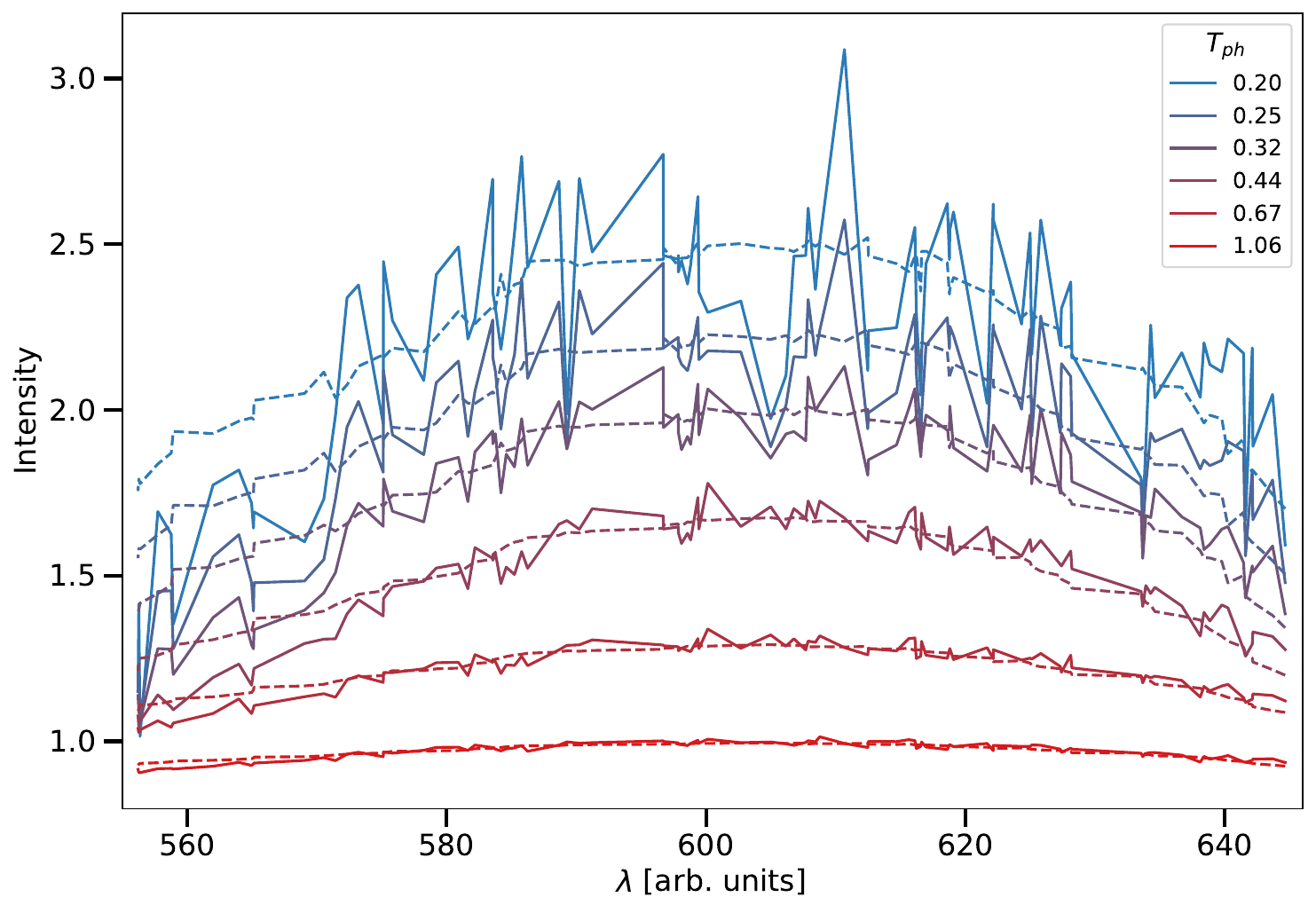}
\caption{Intensity spectrum $I(\lambda)$, for a single realization of the quenched disorder of a smoothed cubic random laser  with $N=96$ modes at different temperatures (from top to bottom) $T=0.2,0.25,0.32,0.44, 0.67,1.06$. Notice the 
narrowing of the central part of the spectrum, because of FMC, as $T$ decreases. The dashed lines are the average spectra over $100$ distinct realizations of the quenced disordered couplings. The values of $\lambda$ are arbitrary and obtained as $\lambda = 1/(\omega_0 + r * (\omega_1-\omega_0))$ with $\omega_0 = 1.55\times 10^8$ and $\Delta \omega = 0.25 \times 10^8$ and $r\in[0,1]$. They are chosen to mimic the spectral domain in nm  of typical random lasers.}
\label{fig:spectrum}
\end{figure}

Given a real material with an emission spectrum extension  between   $\omega_0$ and  $\omega_0+\Delta \omega$,  one can assign the value $\omega_k=\omega_0+r_k*\Delta\omega $ to the frequency.  
For a given fixed instance of the disorder the spectrum $I(\omega_k)\equiv |a_k|^2/\sqrt{T_{\rm ph}}$ in $\lambda = 2\pi c/\omega$ displayed in  Fig. \ref{fig:spectrum}
\footnote{From Eq. (\ref{eq:beta_ph}), $ \beta_{\rm ph}=\epsilon^2/ T =\epsilon^2 \mbox{const}$,
where $T$ is the temperature of the environment of the laser compound, that is fixed, whereas the external pumping ($\propto \epsilon$) changes. We  set  conventionally $T=1$. 
That is, $\epsilon = 1/\sqrt{T_{\rm ph}}$. In the numerical simulations we change $T_{\rm ph}$, the simulation effective temperature, rather than $\epsilon$, the physical pumping rate, and we implement  the smoothed cubic constraint as  $\sum_k |a_k|^4 = N$. 
Starting from the simulated $a_k$, thus,  the real mode amplitude of mode $k$ is $a_k \epsilon^{1/2} = a_k/T_{\rm ph}^{1/4}$ and the  intensity spectrum per mode is $I(\lambda_k)= |a_k|^2/\sqrt{T_{\rm ph}}$.}.

The FMC also introduces non-linear correlations in the interactions
affecting the topology of the interaction network.  Modes with more similar
frequencies are connected by a higher number of quadruplets and,
consequently, they are effectively more coupled. 
As a consequence, as one can see once again in Fig.~\ref{fig:spectrum}, 
modes whose frequencies are at the center of the spectrum  tend to interact more than modes whose frequencies
are at the boundaries. 
Such central band narrowing is akin to the spectra of true
  experimental realizations of random lasers~\cite{Cao99,Cao00}. 

Another relevant
  feature of the intensity spectrum shown in
  Fig.~\ref{fig:spectrum} is that its peaks and fluctuations (with respect to the average spectrum) become more and more pronounced and 
   heterogeneous upon decreasing the temperature.

\subsubsection{Extended mode approximation}
If, as it apparently occurs in some glassy random lasers, modes are spatially extended to wide regions of the optically active compound, then each mode is nonlinearly interacting with very many others. This is called  the ``extended modes approximation'' ~\cite{Zaitsev10}. In this case  the only relevant factor in the design of the mode-coupling network is the FMC, Eq. (\ref{FMC}), rather than spatial confinement of light modes. 

The FMC tends to cut  $O(N)$ interacting
quadruplets with respect to the complete graph, as exactly computed for equispaced frequencies in Ref.~\cite{Marruzzo18}.  
We term $$c_{\rm fc}= \left( \begin{array}{c}
N 
\\ 
4 
\end{array}
\right) = O(N^4)$$
the  total connectivity in the fully connected (complete graph) model and $c_{\rm FMC}$ the total connectivity 
once the FMC is applied to the complete graph. Their ratio
turns out to be
\begin{equation}
\label{Alessia}
\frac{c_{\rm FMC}}{c_{\rm fc}} = \frac{1+2N^2}{N^3}\simeq \frac{2}{N}.
\end{equation}
The total amount of couplings left in the network is $c_{\rm FMC}=O(N^3)$
and each phasor spin in the system will be interacting in $O(N^2)$
quadruplets. Though diluted with respect to the complete graph, the
network is still dense. We refer to it as  {\it complete ML  graph}.

\subsubsection{Random values of the effective coupling constants}
Since the Hamiltonian is an extensive, $O(N)$, energy function,  as we sum over all non-zero contributions in Eq. (\ref{eq:H4}), we have to rescale the couplings with the size $N$. Indeed,  because of the requirement of thermodynamic convergence, the magnitude of each coupling coefficient will be smaller and smaller as the number of modes increases.
If we have a ML  graph, as we discussed above, the number of non-zero terms (i.e., the total connectivity) will grow like $c_{\rm FMC}\sim N^{p-1} = N^3$. 
If the value of the $J$'s is random with zero average, as in the glassy case we are considering here, an extensive Hamiltonian on the ML  graph  is guaranteed by a scaling of  the mean square displacement  like $\sigma_J \sim 1/N^{(p-2)/2} = 1/N$, in the $p=4$ case.

To cope also with more diluted  interaction graphs, we can write a generic connectivity scaling $c\sim N^A$ implying 
\begin{eqnarray}
\sigma_J \sim \sqrt{\frac{1}{N^{A-1}}} \qquad , \qquad A \leq p-1.
\end{eqnarray}

In principle, all couplings involving the same mode will be correlated. However, because of the spatially etherogeneous  nonlinear susceptibility in Eq. (\ref{NonLinearCoup}), and the fact that  each coupling coefficient vanishes as $N$ increases, the role of the correlation will be qualitatively negligible as far as the system displays enough modes. 
For this reason, the couplings will be taken as independent Gaussian random variables in the present work:
\begin{align} \label{Gauss}
\mathcal{P}(J_{k_1 \cdots k_p}) = \frac{1}{\sqrt{2 \pi \sigma_J^2}} \exp\left\{-\frac{J^2_{k_1 \cdots k_p}}{2 \sigma_J^2}\right\} \qquad , \qquad p=4
\end{align}
with  $\sigma_J^2 \sim N^{1-A}$,  to ensure the extensivity of the Hamiltonian and where some obvious rescaling of the coefficients in Eq. (\ref{NonLinearCoup}) has been set \cite{Conti11,Antenucci16,Niedda23a}.

\subsubsection{Graph generation}
Finally, the ML interaction graph over which we run our simulations is generated as follows. First, a virtual complete graph with $c_{\rm fc}$ interactions is generated with ordered quadruplets of indices $k_1<k_2<k_3<k_4$. 
 Then, the FMC filter (\ref{FMC}) is applied to the complete graph. Usually, for the sorted frequencies ($\omega_{k_1}<\omega_{k_2}<\omega_{k_3}<\omega_{k_4}$)
 there will be just one independent permutation  satisfying Eq. (\ref{FMC}), i.e.,  $|\omega_{k_1}-\omega_{k_2}+\omega_{k_4}-\omega_{k_3}|=\gamma$, though we numerically checked  other possible realizations. 
Each time a quadruplet of random frequencies matches Eq.~(\ref{FMC}), the corresponding interaction is added to the real graph and a random value extracted from the Gaussian distribution~\eqref{Gauss} with $p=4$ is assigned to it.

Each one of the $N_{\text{s}}$ disordered samples simulated is characterized by a realization of the couplings $\{J_{\bm{k}}\}$ that differs from the others both for 
the quadruplet networks and for the numerical values.

\subsection{(Avoiding) intensity condensation}
\label{sec:nocond}
\begin{figure}
\includegraphics[width = .99\columnwidth]{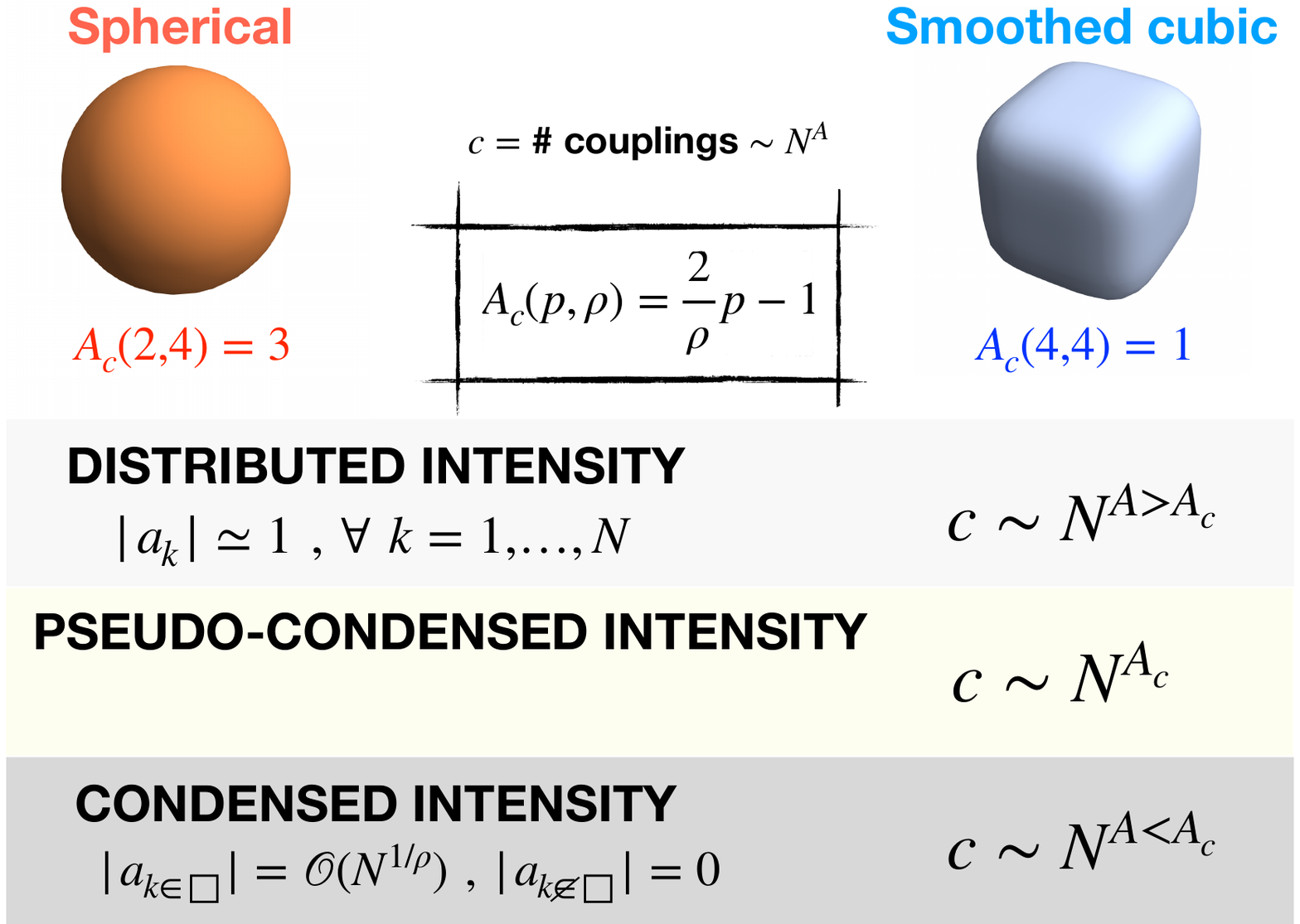}
\caption{Scheme of the network connectivity scaling in relationship with the possibility of condensation. The case studied in the present work is the smoothed cubic one ($\rho=4$) with $p=4$ multi-spin interaction.}
\label{Fig:condensation}
\end{figure}

A constraint (\ref{SmoothedCubic}) stronger than the spherical one, besides being a more realistic approximation of the gain saturation,   also yields a technical advantage. Indeed, any system with interacting continuous variables subject to some local or global potential constraint can undergo {\it intensity condensation} if the interaction network is not dense enough.
In figure \ref{Fig:condensation} we give a  sketch of the possible regimes in the intensity distribution among modes in terms of the order $p$ of multi-spin interaction, the scaling exponent $A$ of the connectivity in the Hamiltonian (\ref{Hamilt2}) and the exponent $\rho$ of the global constraint \eqref{SmoothedCubic} of the mode intensities that directly influences the distribution of the total energy  in the random medium. In appendix \ref{app:condensation} we report the argument leading to the scaling regimes reported in the figure.
The main issue  here is that if the  scaling of the number of non-zero interaction terms is 
\begin{equation}
 \mbox{total \# couplings} \equiv c\sim N^A
\label{eq:con_scaling}  
\end{equation}
there is a critical value $A_c(p,\rho)$ for the scaling exponent  
\begin{equation}
\label{Eq:Ac}
A_c(p,\rho)\equiv \frac{2p}{\rho}-1 ,
\end{equation}
below which the power pumped into the laser condenses on a finite number of modes. 

 In particular, since changing the global constraint from spherical to smoothed cubic $A_c$ drops from $3$ to $1$, no pseudo-condensation is expected to occur in the ML complete graph. This will be checked and reported in Sec. \ref{sec:IPR}, Fig. \ref{eq:yn_scaling_lattice}.
 We also argue that since in the smoothed cubic case intensity condensation is avoided also for $c\sim N$
 it is possible to simulate the model on sparse ML  graphs, in which each mode is connected to a finite number of other modes, independently of their total number $N$. Preliminary simulations confirm such conjecture \cite{BenedettiTesi} and work is in progress to perform simulations  optimized to exploit the sparsness advantage. 
 
Now that the properties of the model have been reported in some detail we can move on to the numerical techniques that we have been using and developing in order to study the dynamical properties of the model at stationarity.

\section{Monte Carlo simulations}

When performing the numerical simulation on a computer, 
the model we are dealing with is very difficult to thermalize to equilibrium and suffers of very strong finite size effects. 
The system, as any spin-glass-like system, is known to be Non-deterministic Polynomial Complete (NPC) \cite{Garey79}. Operatively, to look for equilibrium states, the simulation time needed  scales with the number of modes $N$ approximately as $e^{AN}$. 
Furthermore, our spins are continuous (complex) variables and we have no cunning shortcuts as, for instance, the multi-spin coding that one can exploit for Ising spins, accelerating the computation with bitwise operations, parallelizing different realization of the disorder and allowing to simulate systems of larger sizes in reasonable computing times. 
Finally, the interaction network is {\it dense}, i.e., each spin interacts with a number of other spins that grows with $N$, in particular as $N^2$ in the complete ML graph under probe.

We, therefore resort to several methods to sensitively reduce the equilibration time:  (i) using the Exchange Monte Carlo (EMC) algorithm~\cite{Hukushima96} and parallelizing the simulation of the various copies in temperature   on GPU, (ii) parallelizing the computation of the energy difference of each single proposed spin update on GPU. The system clearly stays NPC, but we have been able to thermalize different sizes deep in the lasing regime. 

We, herewith, describe the main features of the numerical approach, starting from the update proposal of the spin phasors $\{a\}$.

\subsection{Algorithmic sketches}
\subsubsection{Single Metropolis update} 
The single (actually {\it double}) update for spins that are continuous complex and subject to a global constraint is constructed by selecting two phasor spins $a_j$, $a_k$  at random and proposing a random update of both spins $a_j\to a_j'$, $a_k\to a_k'$ which locally preserves their contribution to the global spherical constraint (\ref{SmoothedCubic}): 
each update of the dynamics
must be compatible with the smoothed cubic  constraint $$|a_j|^4+|a_k|^4 = |a_j'|^4+|a_k'|^4
\equiv K^4. $$  Writing the phasors in their polar form, $a_j = A_j e^{\imath \phi_j}$ and $a_k=A_ke^{\imath\phi_k}$, each one of the phases can be updated satisfying such constraint, with the 
 extraction of two random numbers 
$\phi_i', \phi_j'\in [0,2\pi]$.
Then, to modifiy the magnitudes, a third random number $\theta \in [0, \pi/2]$ is extracted. It proposes a move to another point on the surface  of the smoothed-square $A_j^4+A_k^4=K^4$, specifically in the positive coordinates sector, since   $A_i >0$, $\forall i$. The new configuration is 
$$A_j' = K \sqrt{\cos \theta}\quad , \quad A_k' = K \sqrt{\sin \theta}.$$

All together it amounts to extract three pseudo-random numbers for each update proposal
\footnote{The careful reader has certainly noticed that this choice samples uniformly on the quarter of circle rather than on the quarter of smoothed-square. That is, it proposes updates on the latter more frequently on the flat sides and less frequently around the vertex. Though this violates strict detail balance (however, no more than the serial update often used in Monte Carlo simulations), we checked that it does not alter the simulation behavior at equilibrium and it is much faster than uniformly sampling the smoothed square. The latter would, indeed, involve the numerical computation of the arc length along the smoothed square and its inversion.}.

A single Monte Carlo step or {\it sweep} (MCS) is realized by repeating $N/2$ times the update of randomly chosen couples of spins.

\subsubsection{Exchange Monte Carlo - parallel Tempering} 
The numerical simulations have been performed 
by means of an Exchange Monte Carlo algorithm~\cite{Hukushima96} parallelized on GPUs to sample  the equilibrium probability distribution Eq.~\eqref{ProbDistr}. The Exchange Monte Carlo, else called Parallel Tempering (PT) is a widespread tool for thermalizing   ``hardly-relaxing" systems, characterized by  a rugged free energy. It is based on the idea that the thermalization is facilitated by a reversible Markovian dynamics  between system replicas in different heat baths. In particular, configurations belonging to copies of the system at higher temperature help the copies at lower temperature to jump out of minima of the rugged free energy landscape. 
For each size $N$ of the simulated systems, we have run PT simulations with $N_{\text{PT}}$ thermal baths at temperatures $T \in [T_{\rm min}, T_{\rm max}]$. The values are reported in Table \ref{tab1}. 
\begin{table}[t!]
\centering
\begin{tabular}{lrrrrrrr}
\hline \hline
$N$ & $T_{\text{min}}$ & $T_{\text{max}}$ & $N_{\text{PT}}$  & $N_{\text{MCS}}$ & $N_{\text{rep}}$ & $N_{\text{s}}$ \\
\hline 
18 & 0.15 & 1.06 & 42 & $2^{23}$ & 4 & 1000 \\
32 & 0.15 & 1.06 & 42 & $2^{23}$ & 4 & 1000 \\
48 & 0.15 & 1.06 & 42 & $2^{23}$ & 4 & 1000 \\
64 & 0.15 & 1.06 & 42 & $2^{23}$ & 4 & 540 \\
80 & 0.2 & 1.06 & 37 & $2^{23}$ & 2 & 100 \\
96 & 0.2 & 1.06 & 37 & $2^{23}$ & 2 & 100 \\
112 & 0.3 & 1.06 & 28 & $2^{23}$ & 2 & 100 \\
\hline \hline
\end{tabular}
\caption{Details for the simulations of the 4-phasor model with smoothed cubic global constraint on complete ML  graphs.}
\label{tab1}
\end{table}

Each copy at each temperature shares the same realization of quenched disordered couplings $\{J_{\bm{k}}\}$.
The Metropolis dynamics for the spin updates is carried out in parallel in each thermal bath and once each $64$ steps (except for $N=112$ for which $128$ MCS are taken) an exchange of configurations, a {\it swap} between baths at neighbouring temperatures is proposed. A swap is proposed sequentially for all pairs of neighbouring inverse temperatures $\beta_i$ and $\beta_{i+1}$, with the following acceptance probability, implementing detailed balance with the equilibrium Boltzmann distribution for each thermal bath:
\begin{align}
p_{\text{swap}} = \min\ [1 \ ,\ e^{(\beta_i - \beta_{i+1})( \mathcal{H}[\bm{a}_i] -  \mathcal{H}[\bm{a}_{i+1}])}].
\end{align}
For all simulations the $N_{\text{PT}}$ temperatures have been taken with an adapting spacing in $T$, to optimize the  equilibration of the system down to low $T$, see App. \ref{app:deltaT}.

 In between swaps, each copy in the various thermal baths is simulated in parallel on GPUs.

\subsubsection{Energy update parallel computation} 
The computation of the energy difference, in the Metropolis update, between configurations of complex continuous spins is an essential feature. 
The energy shift between the original configuration 
and one in which one spin is changed, $a_\ell \to a'_\ell$, requires a number of operations  scaling with
 the number of quadruplets involved in the variation of $a_k$
\begin{eqnarray}
\nonumber
    \Delta E_k &=& \left(a_\ell-a_\ell'\right)
    \hspace*{-.5cm}\sum_{k_1,k_2,k_3}^ {{\rm FMC}(k_1,k_2,k_3,\ell)}\hspace*{-.5cm} J_{k_1k_2k_3\ell}\  \bar a_{k_1} a_{k_2} \bar a_{k_3} + \mbox{c.c.} 
    \\
    \nonumber
    &=& \mathcal{O}(N^2),
\end{eqnarray}
since our random laser model on a complete ML  graph has a connectivity per node growing like $N^2$. Therefore, the time it takes to accept or refuse the update yielding 
an energy variation $\Delta E = \mathcal H[\bm a']-\mathcal H[\bm a]$, after a spin-couple update $a_j\to a_j'$, $a_k\to a_k'$ grows like  $O(N^2)$, as well. Implying a scaling with $N^3$ of a whole MCS ($N/2$ updates of spin couples).
To cope with this bottleneck, in our code   the energy contribution of each quadruplet to   $\Delta E$   is computed apart, on parallel kernels on GPU and further summed in parallel using a number of  operations scaling like $O(\log N)$. 
Operatively, since the number of parallel threads exceeds the number of cores in the GPU, the complete parallelization is not feasible. Indeed, the actual scaling with $N$ of an entire MCS  is still growing like $N^3$, but with a prefactor of orders of magnitude smaller than with a serial $\Delta E$ computation (approximately $0.166(4)$ seconds per $N^3$).
The code, written in CUDA, has been running on  Nvidia Tesla V100 (5120 cores).

\subsection{Numerical tests}
\subsubsection{Data thermalization}
All observables analyzed in the following are drawn from configurations at equilibrium. 
To test thermalization we used two methods. 

First we looked at energy relaxation on sequential time windows whose length is  double each time with respect to the previous one. 
As a further test we check the symmetry of the overlap distribution $P_J(q)$ - the order parameter of the glass transition, - for single samples of the bond disorder.
All relevant observable averages, including $P_J(q)$, are computed on such exponentially growing time scales, in order to compare their values on different simulation time scales to identify the equilibrium values for each apart disordered sample. In case the simulation time is long enough to reach equilibrium average over the last time window (half of the simulation time) are reported.

Once thermalization to equilibrium dynamics has been tested and a thermalization time $\tau_{\rm eq}$ identified, the time average coincides with the canonical ensemble average. In order to provide the comparable statistics at all temperatures, if a simulation of $N_{\rm MCS}$ Monte Carlo steps thermalizes at all temperatures, we take $\tau_{\rm eq} = N_{\rm MCS}/2$. 

\subsubsection{Data time correlation}
 In order to properly estimate statistical errors, time correlations have been taken into account, as well. A correlation time $\tau_{\rm corr}$  has been identified as the maximum among all thermal bath dynamics, approximately equal to $256=2^8$ Monte Carlo steps. 
 Consequently, we measure the observables every $\tau_{\rm corr}$ Monte Carlo steps.
If $N_{\rm MCS}$ is the total amount of Monte Carlo steps of the simulation, for each disordered sample we, thus, record 
\begin{equation}
    \mathcal N \equiv \frac{N_{\rm MCS}-\tau_{\rm eq}}{\tau_{\rm corr}}
\label{eq:statistics}    
\end{equation}
thermalized, 
uncorrelated configurations $\bm a_t$.

\subsubsection{Averages and errors}
For a single sample of the coupling disorder $\{J_{\bm{k}}\}$, cf. Eq. (\ref{Gauss}),
the ensemble average unbiased estimate is 
\begin{align}
\label{eq:totave}
\langle O[\bm{a}] \rangle_J = \frac{1}{ \mathcal N}\sum_{t=\tau_{\rm eq}/\tau_{\rm corr}}^{N_{\rm MCS}/\tau_{\rm corr}} O[\bm{a}_t]. 
\end{align}
For a random couplings realization labeled by $\mu$ we have a thermal average $\langle O[\bm{a}] \rangle_J^{(\mu)}$.
In our analysis, we repeat the simulations on $N_s$ different random samples of randomly determined complete ML  graphs in which the values of the quadruplet coupling constants are extracted by means of Eq. (\ref{Gauss}). 

  Averaging over the random samples $\mu=1,\ldots, N_s$ yields the least fluctuating finite $N$ proxy for the average in the thermodynamic limit:
\begin{align} \label{DisAver}
\overline{O} = \frac{1}{N_{\text{s}}} \sum_{\mu=1}^{N_{\text{s}}}  \langle O[\bm{a}] \rangle_J^{(\mu)}.
\end{align}

Whenever necessary, statistical errors are computed with the jackknife method  in order to reduce the distortions one would have applying error propagation \cite{Miller64,Young15,Leuzzi2025}.

\section{Observables and results}
\subsection{Critical behaviour for dense mean-field models}

Let us consider first a  statistical mechanical model in which a linear size $L$ and a  length are naturally defined (like a lattice of any geometry, for instance a $d$-dimensional cubic lattice) and that undergoes a continuous phase transition at some critical temperature $T_c$.
If the size of the system is $N$, it will be $L=N^{1/d}$. 
In finite size systems there is no actual phase transition, but one can approximate the critical point by a  finite-size critical temperature $T_c(N)$ (we will see how in a while).
Once different $T_c(N)$ are estimated at different sizes $N$ the critical temperature in the thermodynamic limit can be extrapolated as  $T_c = \lim_{N \to\infty} T_c(N)$.

Let us introduce the reduced temperature $t\equiv |T/T_c - 1|$ and its finite $N$ proxy $t_N = |T/T_c(N) - 1|$.
According to the theory of critical phenomena, at criticality the correlation length will diverge as $\xi\sim t^{-\nu}$, the local susceptibility as $\chi \sim t^{-\gamma}$ and, in general, any relevant observable will diverge as $O\sim t^{-\psi_O}$. The exponents $\nu, \gamma, \psi_O $ are the critical exponents of the system and they identify its universality class. 
In this list one must include the critical exponent $\beta$ of the growth of the order parameter next to the critical point: $q\sim 
t^\beta$, $t<0$.

For the observable $O$, near the critical point, the following scaling behavior with $N$ will hold:
\begin{eqnarray}
    \label{eq:o_scaling_lattice}
    O_N(T) = N^{\frac{\psi_O}{\nu d}} \hat f_{O}\left(
    N^{\frac{1}{\nu d}} t_N
    \right)
\end{eqnarray}
where 
$\hat{f}_O$ is a  dimensionless function that depends on the specific observable.
By considering the hyperscaling relation $\nu d =  2 \beta + \gamma$, derived 
matching order-parameter and susceptibility fluctuations on the two sides of the transition\cite{Cardy88} we can rewrite the above equation as 
\begin{eqnarray}
    \label{eq:o_scaling}
    O_N(T) = N^{\frac{\psi_O}{2 \beta + \gamma}} \hat f_{O}\left(
    N^{\frac{1}{2 \beta + \gamma}} t_N
    \right) . 
\end{eqnarray}

In dense mean-field systems without a natural spatial dimension, like our model, there is no geometric critical length scale. The only ``size'' of the system is the number of modes  $N$.
As a consequence, finite-size effects can only be described in terms of how the effective correlation volume grows with 
$N$, since no length is there. 
 To this purpose an effective exponent is introduced,  
  related to the critical exponent $\beta$ of the order parameter and the critical exponent $\gamma$ of the susceptibility, well defined also when a characteristic length is not. It is  
  \begin{equation}
  \label{def:nueff}
      \nu_{\rm eff}= 2 \beta+\gamma.
  \end{equation}
  In mean-field models this combination controls the width of the critical region in finite-size scaling \cite{Niedda23a}.
  
\subsection{Specific heat}
\begin{figure}[t!]
    \centering
    \includegraphics[width=0.99\linewidth]{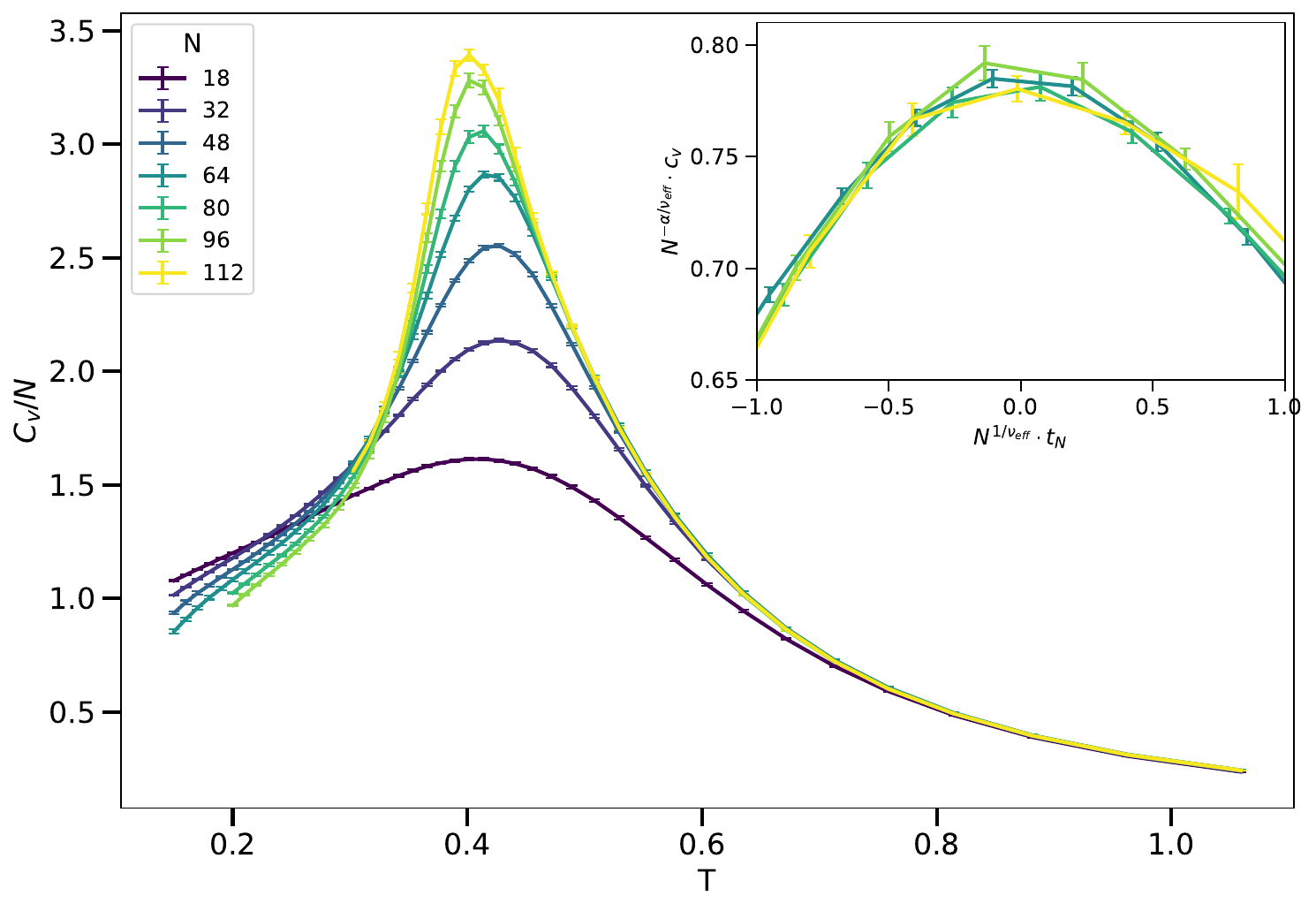}
    \caption{Specific heat per spin of the $4$-phasor model on complete ML graph. for sizes $N\in[18,112]$. In the inset  $c_V/N^{\alpha/\nu_{\rm eff}}$ is displayed for the largest sizes vs the rescaled reduced temperature, showing the collapse along a unique curve $\hat f_{C_V}(N^{1/\nu_{\rm eff}}t_N)$, cf. Eq. (\ref{eq:cv_scaling}). }
    \label{fig:cv}
\end{figure}

According to the prescription (\ref{eq:totave}),
at given size $N$ the specific heat per spin $C_V$ is
measured by calculating the energy fluctuations at on data at equilibrium at each
temperature $T$ and then averaging over the disordered samples 
\begin{align}
\label{SpecificHeat}
C_{V} =  \frac{\overline{\langle \left(E - \langle E \rangle_J\right)^2 \rangle_J}}{T^2}, 
\end{align}
where, once again, $\langle\dots\rangle_J$ represents the thermal average at fixed coupling realization $\{J_{\bm k}\}$ and
$\overline{[\dots]}$ represents the average over disorder.
The temperature behavior of the specific heat per spin, $C_V/N$, is displayed in Fig. \ref{fig:cv} for six sizes ranging from $18$ to $112$. 
One can observe that increasing the size the curve is more and more peaked around a size dependent temperature, that is a proxy for the finite size critical temperature $T_c(N)$. 

Next to criticality, according to Eqs. (\ref{eq:o_scaling},\ref{def:nueff}), the specific heat behaves like
\begin{align} \label{eq:cv_scaling}
c_V \equiv \frac{C_{V}(T)}{N} = N^{\frac{\alpha}{\nu_{\rm eff}}} \hat{f}_{C_V}\left(N^{\frac{1}{\nu_{\rm eff}}} ~t_N\right),
\end{align}
where 
the exponent $\alpha$ is the critical exponent of the specific heat peak
divergence at criticality, $c_V \sim t^{-\alpha}$, $t\equiv |T/T_c-1|$. 


Now, since the adimensional function  $ \hat{f}$ is scaling invariant,
if one uses the correct values of the
exponents $\alpha$ and $\nu_{\rm eff}$, the rescaled curves $c_{V}(T)/N^{\alpha/\nu_{\rm eff}}$  should collapse on the same curve for
different values of $N$ in the critical region $t\sim 0$, as displayed in the inset of Fig. \ref{fig:cv}.

\begin{figure}[t!]
    \centering
\includegraphics[width=0.79\linewidth]{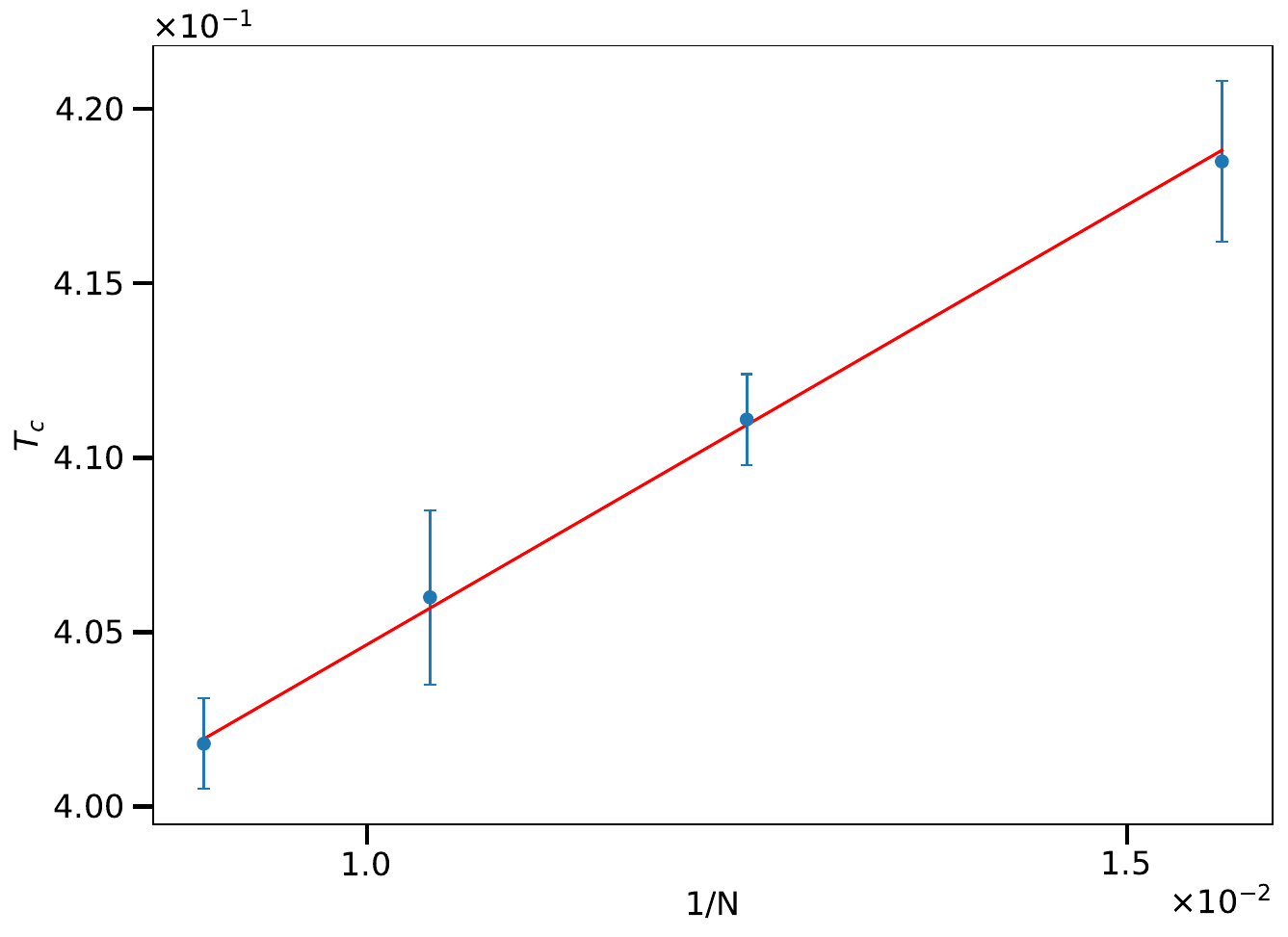}
    \caption{Finite size behavior of the estimate of the specific heat peaks, $T_c(N)$. Finite size scaling Eq. (\ref{eq:FSS}) has been performed including the $4$ largest sizes, on order to avoid pre-asymptotic effects from too smal $N$.  }
    \label{fig:CvTc}
\end{figure}

In order to assess the  $T_c(N)$ we fit the points around the peak of each $c_V$ curve in Fig.~\ref{fig:cv}, with a quadratic function of the temperature $f_N(T)=a_N+b_NT+c_NT^2$ and compute the maximum point of each fitting function as $T_c(N)=-b_N/(2c_N)$, estimating the statistical error accordingly. 
The critical temperature $T_c(\infty)$ of the model can be extrapolated from the fit of the finite-size critical temperatures, cf. Fig. \ref{fig:CvTc}, with the following function: 
\begin{equation}
T_c(N) = T_c(\infty) + a N^{-1},
\label{eq:FSS}
\end{equation}
Neglecting the smallest sizes to avoid preasymptotic effects, see Fig. \ref{fig:CvTc}, we obtain

\begin{equation}
T_c = 0.379 \pm  0.004
    \label{eq:Tc}
    \end{equation}

\begin{figure}[t!]
    \centering
    \includegraphics[width=0.51\linewidth]{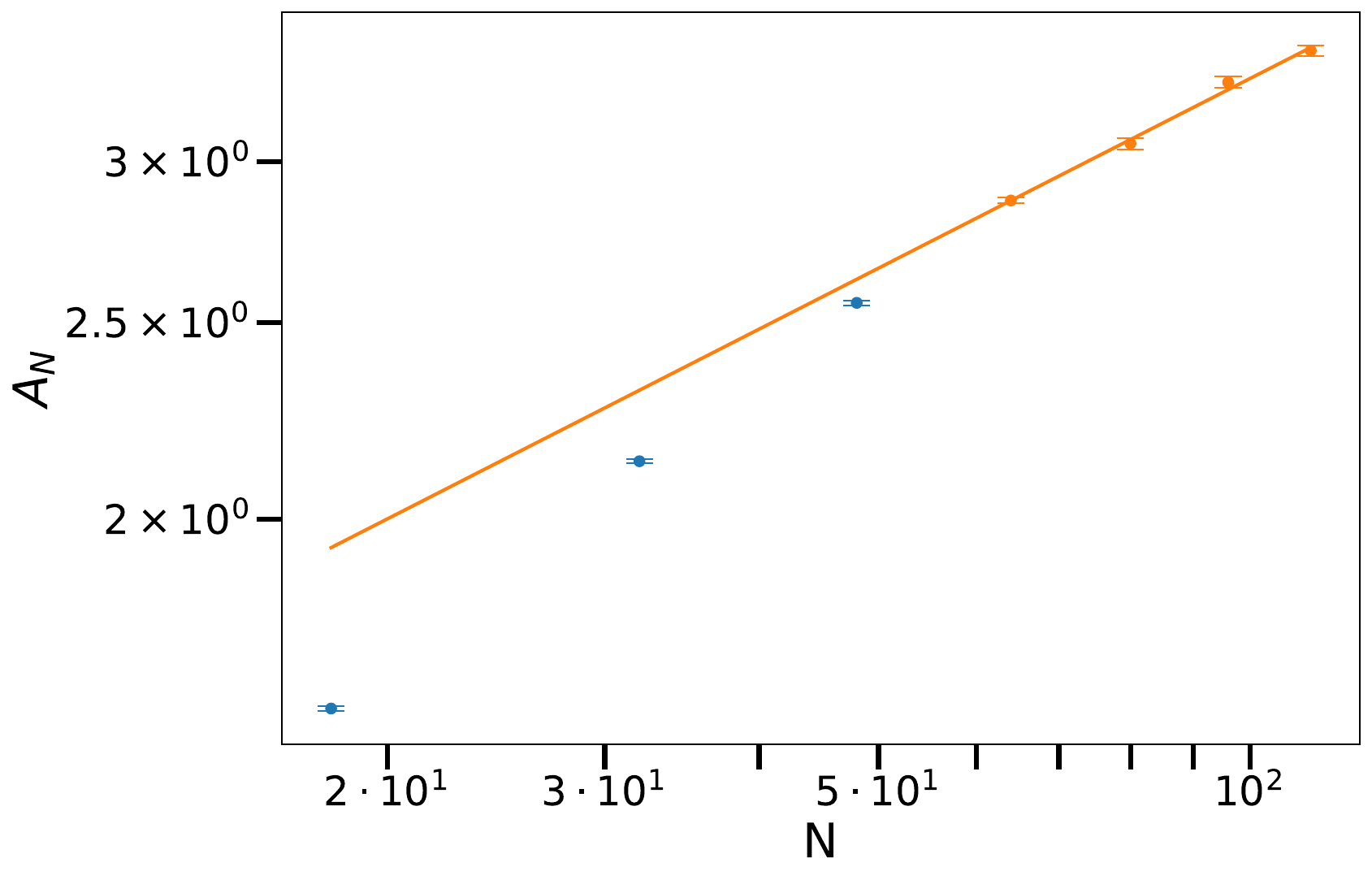}
    \includegraphics[width=0.465\linewidth]{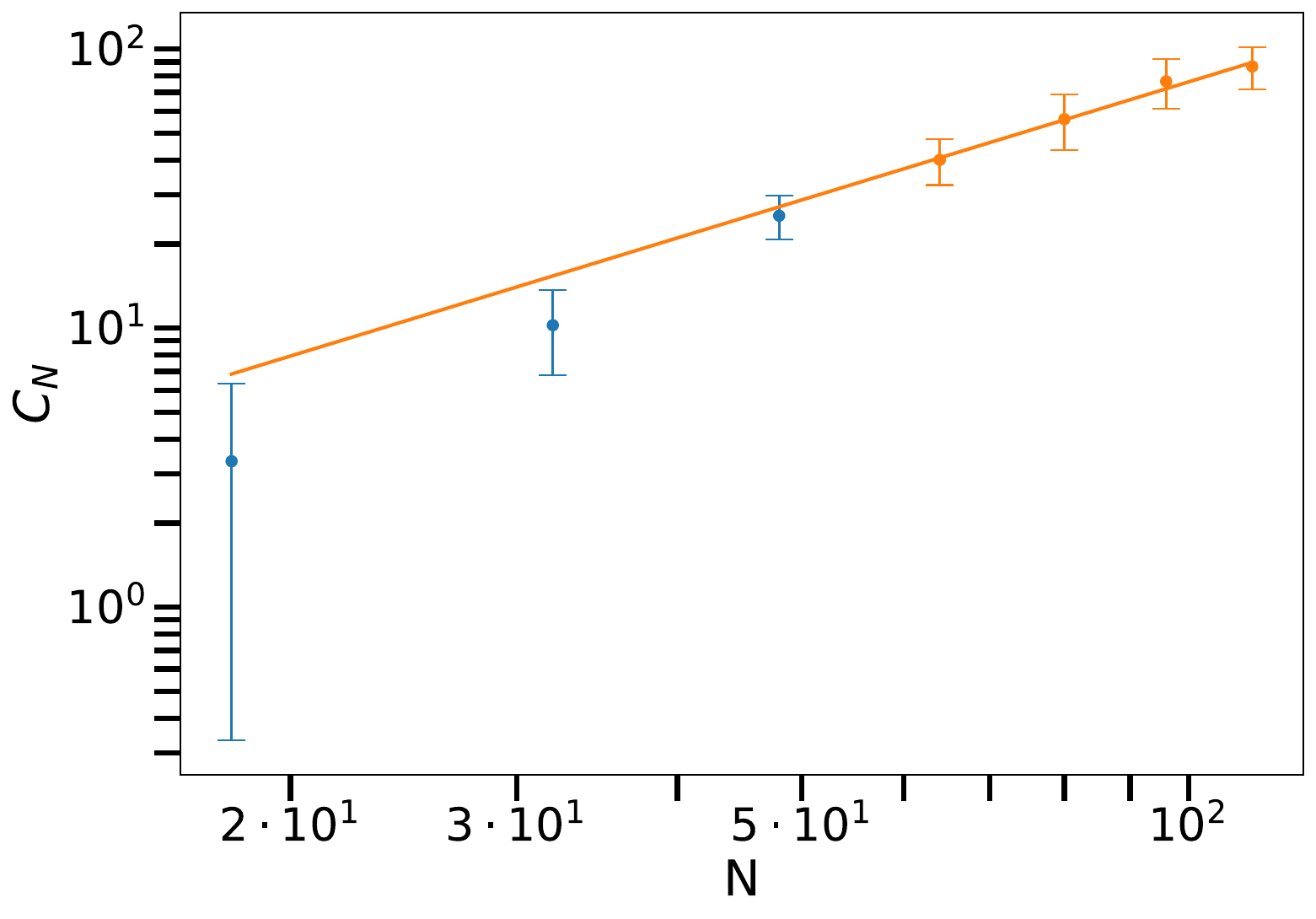}
    \caption{$A_N$, $C_N$ estimates from finite size scaling of the peaks of the specific heat. The interpolation as been carried out using only the four larger sizes. }
    \label{fig:crit_cv}
\end{figure}

We then take the following Ansatz on the form of scaling function $\hat{f}_{C_V}$ in Eq.~\eqref{eq:cv_scaling}
\begin{align}
\hat{f}_{C_V}(x) = A + C x^2,
\end{align}
where $x=N^{1/\nu_{\rm eff}} t_N$.
With this Ansatz the scaling hypothesis for the specific heat Eq.~\eqref{eq:cv_scaling} reads as
\begin{align}
c_{V_N}(T) = \tilde{A}_N + \tilde{C}_N t_N^2
\end{align}
where $\tilde{A}_N = A_N N^{\frac{\alpha}{\nu_{\rm eff}}}$ and $\tilde{C}_N = C_N N^{\frac{\alpha + 2}{\nu_{\rm eff}}}$. Interpolating the points of the $c_V$ curves around their peaks with the previous function we determine the values of the coefficients at each $N$. The behaviour of the logarithm of the coefficients is
\begin{eqnarray*}
\ln \tilde A_N &=& \ln A_N +\frac{\alpha}{\nu_{\rm eff}}\ln N,
\\
\ln |\tilde C_N| &=& \ln |C_N| +\frac{\alpha+2}{\nu_{\rm eff}}\ln N
\end{eqnarray*}
we, thus, estimate $\alpha$ and $\nu_{\rm eff}$ by linear interpolation in $\ln N$.
The finite size scaling (FSS) analyses is reported in Fig. \ref{fig:crit_cv}
and yields the values
\begin{eqnarray}
    \label{eq:alpha}
    \alpha =  0.57\pm 0.23 \quad , \quad  
\nu_{\rm eff}= 1.83\pm 0.72.
\end{eqnarray}  
These estimates of $\alpha$ and $\nu_{\rm eff}$  are compatible with the values of the critical exponents for Random Energy Model, which is the
reference mean-field model for disordered systems with non-linear interactions \cite{Derrida80,Derrida81,Mezard84,Mezard09}. In that case $\nu_{\rm eff}=2$ is expected and numerical simulations on REM provided \cite{Niedda23a}: $\alpha = 0.52 \pm 0.07$ and $\nu_{\rm eff} = 1.94 \pm 0.22$.
In general, the range of values for $\nu_{\rm eff}$ for any mean-field-like universality class is $\nu_{\rm eff}\in[1,2]$ \cite{Niedda23a}.

The critical indices of the version of the  $4$-phasor model on the complete ML graph with spherical (other than smoothed-cubic) global constraint and comb-like (other than continuous uniform) frequencies, are, instead  $\alpha = 0.48 \pm 0.05$ and  $\nu_{\rm eff} = 0.91 \pm 0.15$ \cite{Niedda23a}. 
The discrepancy in the $\nu_{\rm eff}$ value might be due to the larger sizes simulated in the smoothed cubic case and, possibly, to smaller finite size effects in the model.

\subsection{Parisi overlap distribution}

Overlaps are defined as scalar products among phasor configurations of independent replicas of the system with the same quenched disorder. In the present case the relevant overlap for the transition turns out to be \cite{Leuzzi09a,Antenucci15e, Antenucci16}
\begin{equation}
q_{cd} = \frac{1}{N} \mbox{Re } \sum_{i=1}^N \overline{a}_k^{(c)} a_k^{(d)} 
\end{equation}
where $c$ and $d$ are replica indexes. Since replica overlaps measure the similarity between glassy states of the system, their distribution gives information about the structure of the phase space.

The order parameter for the glass transition is the overlap probability distribution $P(q)$~\cite{Mezard87}. In models with continuous variables, the $P(q)$ is expected to be a distribution with a single peak in $q=0$ in the high temperature phase and to develop side peaks, as well, in the low temperature glass phase. Of course, at finite $N$,  no exact Dirac delta distribution occurs because of finite size effects, yet smoothed peaks and large tails appearing in  the $P(q)$   yield indications of a non-trivial behavior below a critical point in temperature. 

The protocol used in numerical simulations to measure the overlaps corresponds to the definition of replicas as independent copies of the system with the same quenched disorder. 
For each sample, i.e., each realization of disorder $\{J_{\bm k}\}$, 
we run dynamics independently for $N_\text{rep}$ replicas of the system, starting from randomly chosen initial phasor-spin configurations. In this way, replicas explore different regions of the same phase space and may thermalize in configurations belonging to apart states, if many apart states are there. To study the  behavior of the $P_J(q)$ for the smallest sizes we choose $N_\text{rep}=4$, so that at any measurement time six values of the overlap are available $q_{\alpha\beta}=\{q_{01}, q_{02}, q_{03}, q_{12}, q_{13}, q_{23}\}$, see Tab. \ref{tab1}. We used $N_{\rm rep}=2$ for large $N\geq 80$.
Given a number $\mathcal N$, cf. Eq. (\ref{eq:statistics}, of well thermalized, independent data,
for each disordered sample the $P_J(q)$ histograms are, thus, built with $\mathcal{N} \times  N_\text{rep}(N_\text{rep}-1)/2 $ values of the overlap. 

\begin{figure}[t!]
    \centering
    \includegraphics[width=0.99\linewidth]{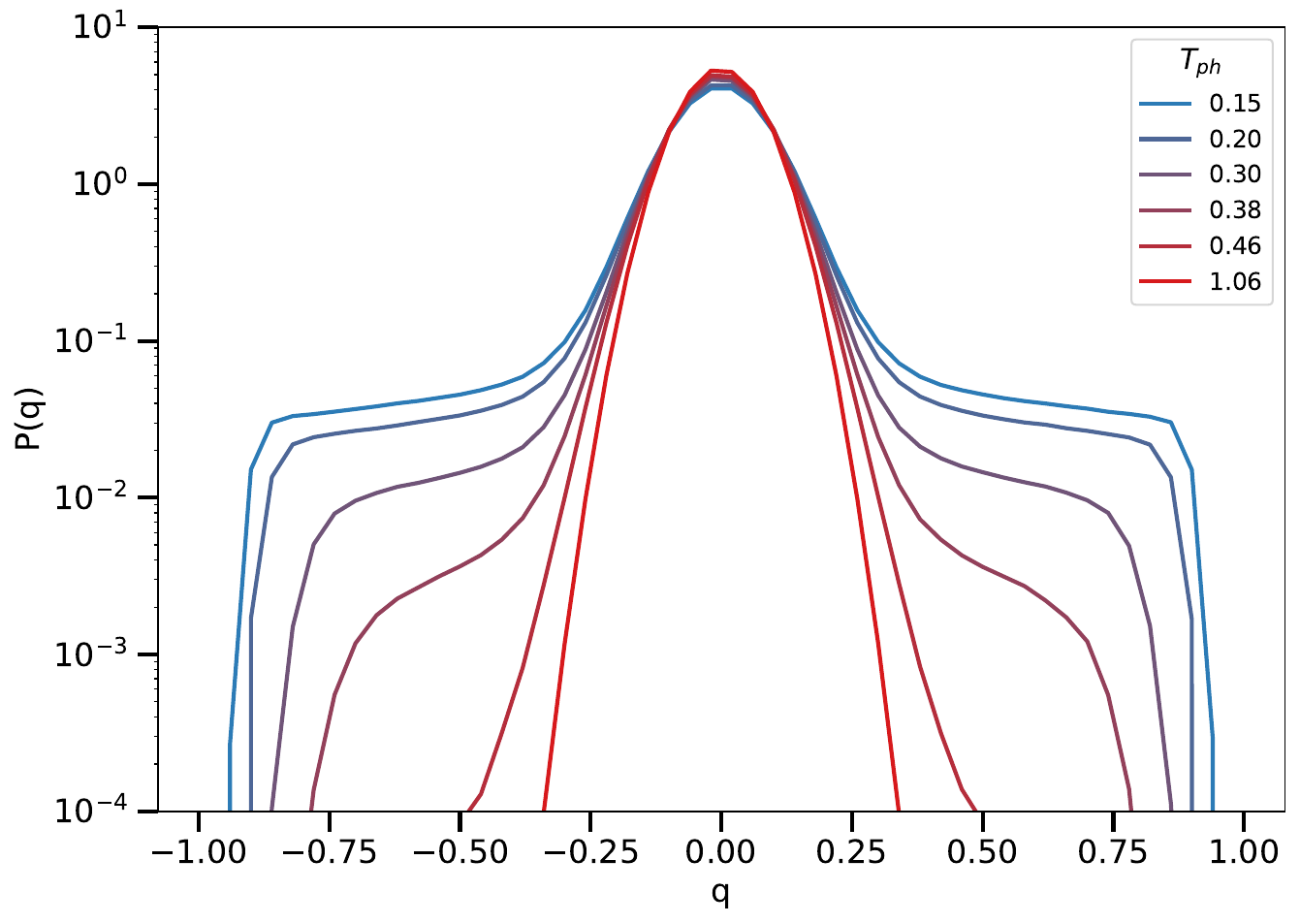}
    \caption{Parisi overlap distributions at $N=64$ from high $T=1.06$ to low $T=0.15$ across the transition point in semi-log scale. }
    \label{fig:pqN}
\end{figure}

\begin{figure}[t!]
    \centering
    \includegraphics[width=0.99\linewidth]{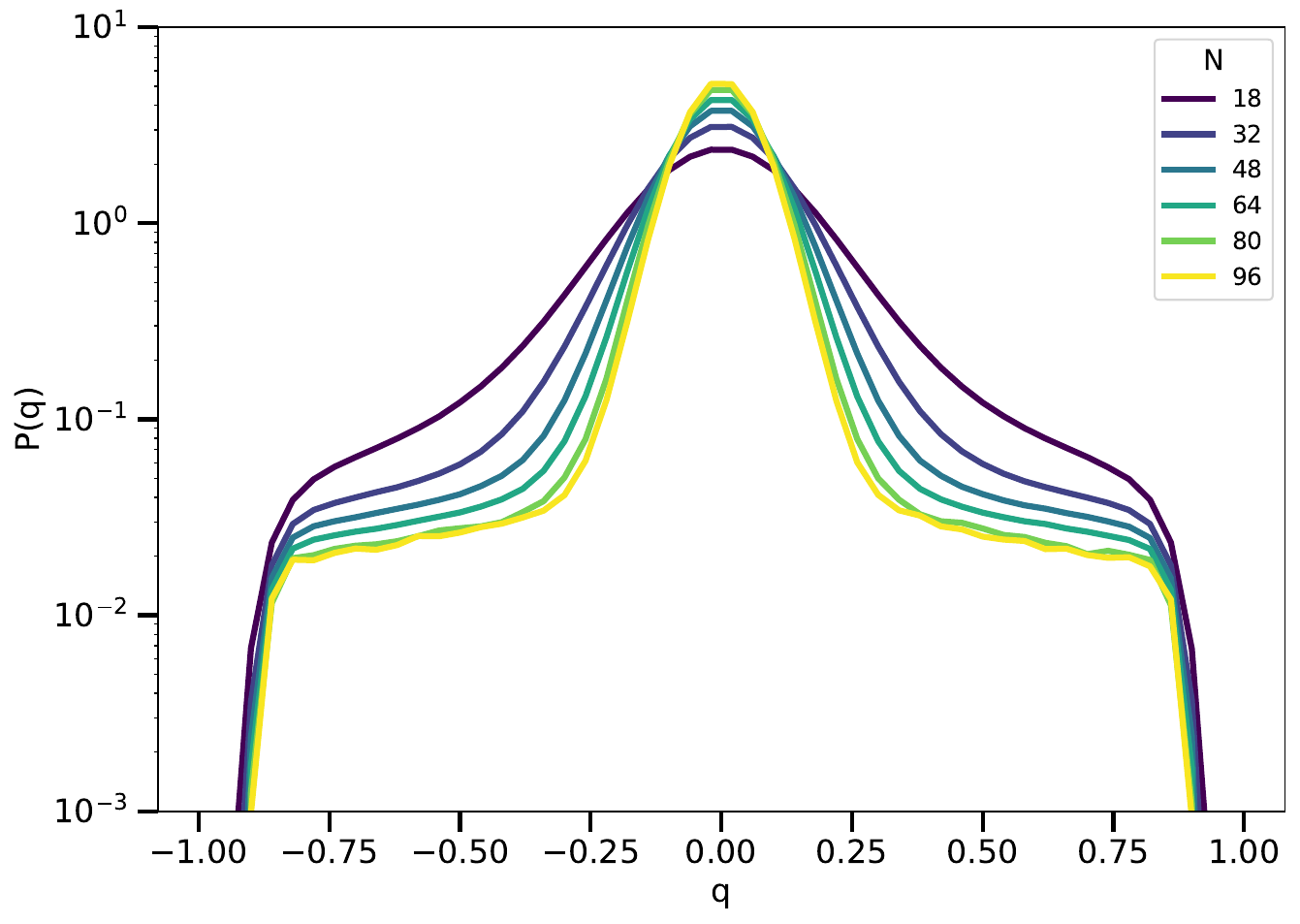}
    \caption{Parisi overlap distributions at  $T=0.2<T_c$ as the size varies.}
    \label{fig:pqT}
\end{figure}

The overlap distribution functions $P_J(q)$ are computed as the normalized histograms of the overlaps for each one of the samples. 
In Figs.~\ref{fig:pqN}, \ref{fig:pqT} we present the average overlap distributions over $N_s$ disordered samples, cf. Tab. \ref{tab1}. 
In Fig. \ref{fig:pqN} we display the behaviour of the  average $P(q)$  in a system of $N=64$ spins at various temperatures. At $T=1.06$ we are in the paramagnetic phase and the $P(q)$ is a Gaussian. As $T$ decreases, at $T=0.46$ we notice that $P(q)$ starts deviating from a Gaussian and for lower $T=0.38$ it starts developing side branches. 

We know that in the exactly solvable spherical mean-field fully connected models in the narrow band approximation the $4$-phasor model displays a so-called {\it random first order} phase transition  to a spin-glass with one breaking of replica symmetry~\cite{Antenucci15a,Antenucci15b,Antenucci15f}. This is represented by an overlap distribution  displaying one central peak and two side peaks on what is termed self-overlap, or Edwards-Anderson overlap $q_{\rm ea}$: $$P(q) = \left(1-\frac{p}{2}\right) \delta(q-q_{\rm ea})
+ p \delta(q) +\left(1-\frac{p}{2}\right)  \delta(q+q_{\rm ea}).$$
This might be the outcome at the thermodynamic limit in the present model as well, with non-trivial mode-locking (network dilution), rather than narrow-band (complete graph) and with constraint (\ref{SmoothedCubic}) with $\rho=4$, rather than $\rho=2$. 
In Fig. \ref{fig:pqT} we display the finite size behavior of the $P(q)$ at  $T=0.2$, the lowest $T$ at which we were able to thermalize size $N=96$.
One can observe that we are still very far away from the  three delta peaks of the analytic fully connected spherical model. The picture displayed is compatible with strong finite size effects and/or a $T$ that is not deep enough into the spin-glass phase, though certainly no final claim can be made about  the thermodynamic limit of the $P(q)$.


To quantitatively estimate the deviation from gaussianity of the $P(q)$ as temperature is lowered we employ the kurtosys
\begin{equation}
    \label{eq:kurt}
    K_N(T) \equiv \frac{{\overline{\left\langle \left(q-{\overline{\langle q\rangle}_J}\right) ^4\right\rangle_J}}}{\left({\overline{\left\langle \left(q-\langle q\rangle_J\right)^2 \right\rangle_J}}\right)^2}.
\end{equation}

This particular observable has the property of being scale invariant at the critical point~\cite{itzykson89b}, 
i.e., 
\begin{align} \label{eq:K_scaling}
K_N  = \hat{f}_K\left(N^{\frac{1}{\nu_{\rm eff}}} ~t_N\right).
\end{align}
This means that at $t_N=0$, i.e., at $T=T_c(N)$ the  $K_N(T)$ curves cross. As far as $T_c(N)$ sensibly shifts with $N$, curves for adjacent $N$ values will cross in (hopefully slightly) different points.

\begin{figure}[t!]
\includegraphics[width=0.49\textwidth]{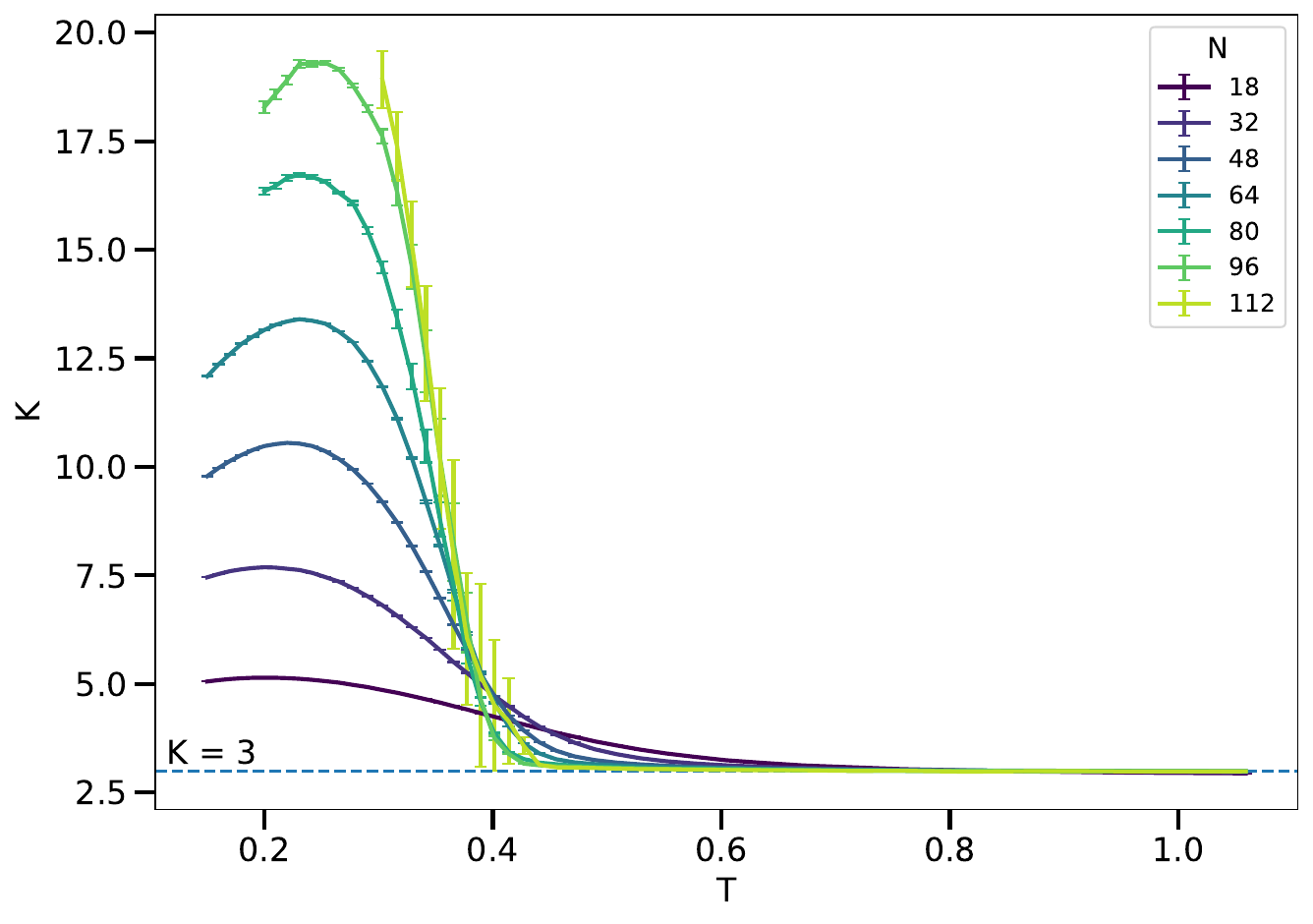}
\caption{Kurtosys of the overlap distribution vs $T$ for $N\in[18,96]$.}
\label{fig:Kurtosys}
\end{figure}

\begin{figure}[t!]
\includegraphics[width=0.79\linewidth]{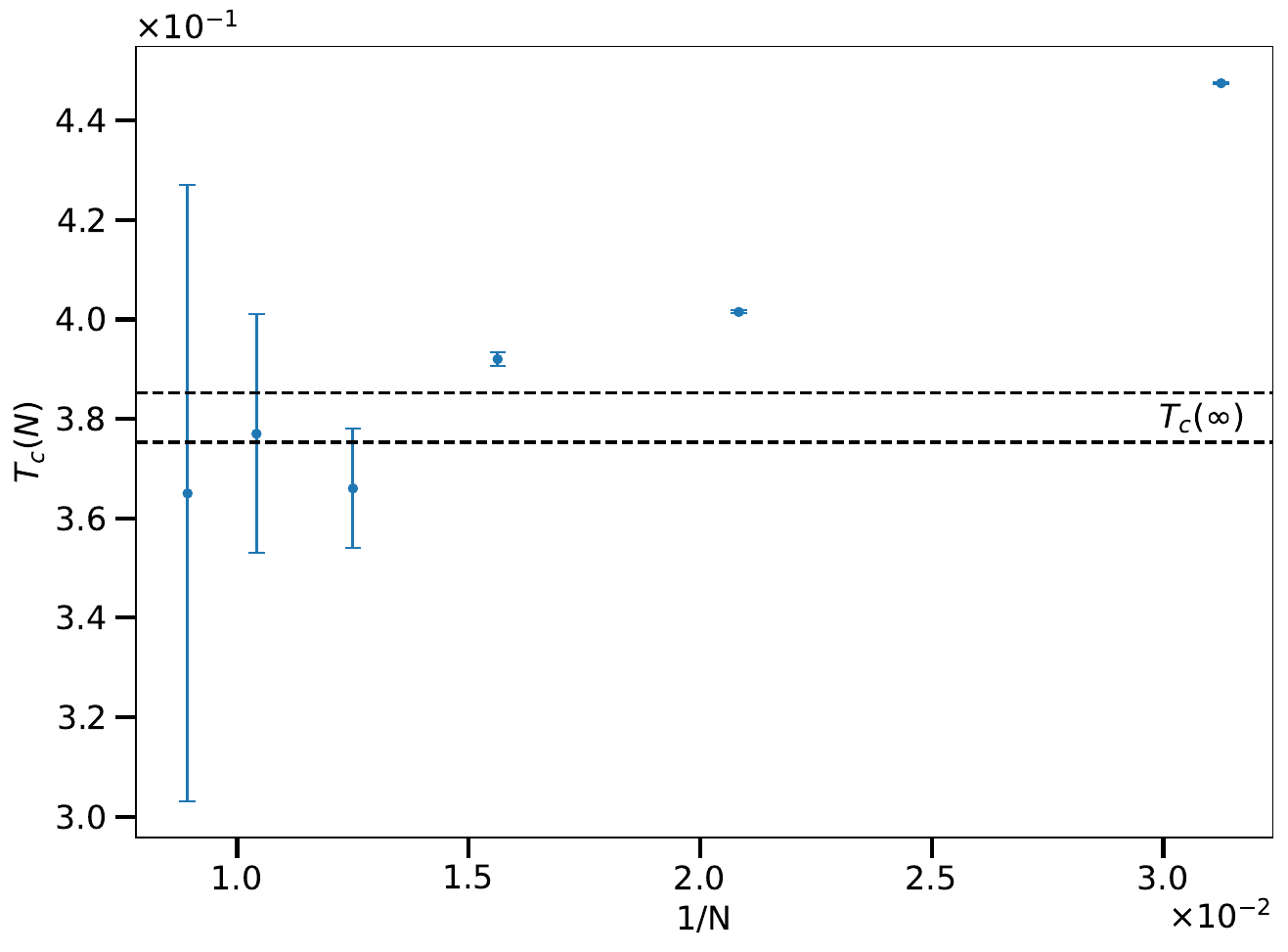}
\caption{Crossing points in temperature of the kurtosys curves $K_N(T)$. The dashed lines are the confidence interval of the specific heat estimate of the critical temperature (\ref{eq:Tc}). }
\label{fig:CrossKurtosys}
\end{figure}

In Fig. \ref{fig:Kurtosys} we display the behavior of the kurtosys vs. $T$ for the simulated sizes. At high $T$ it is $K=3$, as expected for a Gaussian distribution, and lowering $T$ the various curves cross (in different points) around $T\sim 0.4$.   
In Fig. \ref{fig:CrossKurtosys} the crossing points between $K_{18,32}$, $K_{32,48}$, $\ldots$, $K_{80,96}$ are reported with their errorbars vs $1/N$, where $N$ is operatively chosen as the largest one of the crossing curves. A clean finite size scaling of the kind (\ref{eq:FSS}) has not been possible, yet the  data
are compatible with the critical temperature estimate obtained from the specific heat FSS, whose interval is also displayed in the figure with two dashed lines around $T_c(\infty)$, cf Eq. (\ref{eq:Tc}).

\subsection{Inverse participation ratio's}
\label{sec:IPR}

As we mentioned in Sec. \ref{sec:nocond}, a smoothed hyper-cubic  global constraint over the magnitude of the mode intensities should prevent any condensation of the overall energy pumped into the system onto a small (size-independent) number of modes. 
In particular, in the network setting we are considering in this work -- the complete ML graph -- the spherical random laser model is known to display pseudo-condensation \cite{Niedda23b}. That is, intensity is still distributed among all modes but a $O(1)$ number of modes acquire a finite fraction, $O(N)$, of the total intensity. The explanation for this is that the complete ML graph, having a scaling exponent of the number of couplings $A=A_c(4,2)=3$, cf. Eq. (\ref{Eq:Ac}) is at the border between intensity equipartition and condensation.

The quantitative parameters to interpret the behaviour of the system in terms of (lack of) are the inverse participation ratio's (IPR):
\begin{equation}
    Y_n\equiv \frac{\sum_{k=1}^N |a_k|^{2n}}{\left(\sum_{k=1}^N |a_k|^{n}\right)^2 } =  \frac{\sum_{k=1}^N I_k^{n}}{\left(\sum_{k=1}^N I_k^{n/2}\right)^2 }.
\end{equation}
Specifically, in our smoothed cubic case $n=2,4$ will be useful. 

Let us see what  information about condensation we can discriminate through the PR's.  We recall that $\sum_k |a_k|^4=N$ in our model.

\begin{figure}[t!]
    \centering
    \includegraphics[width=0.99\linewidth]{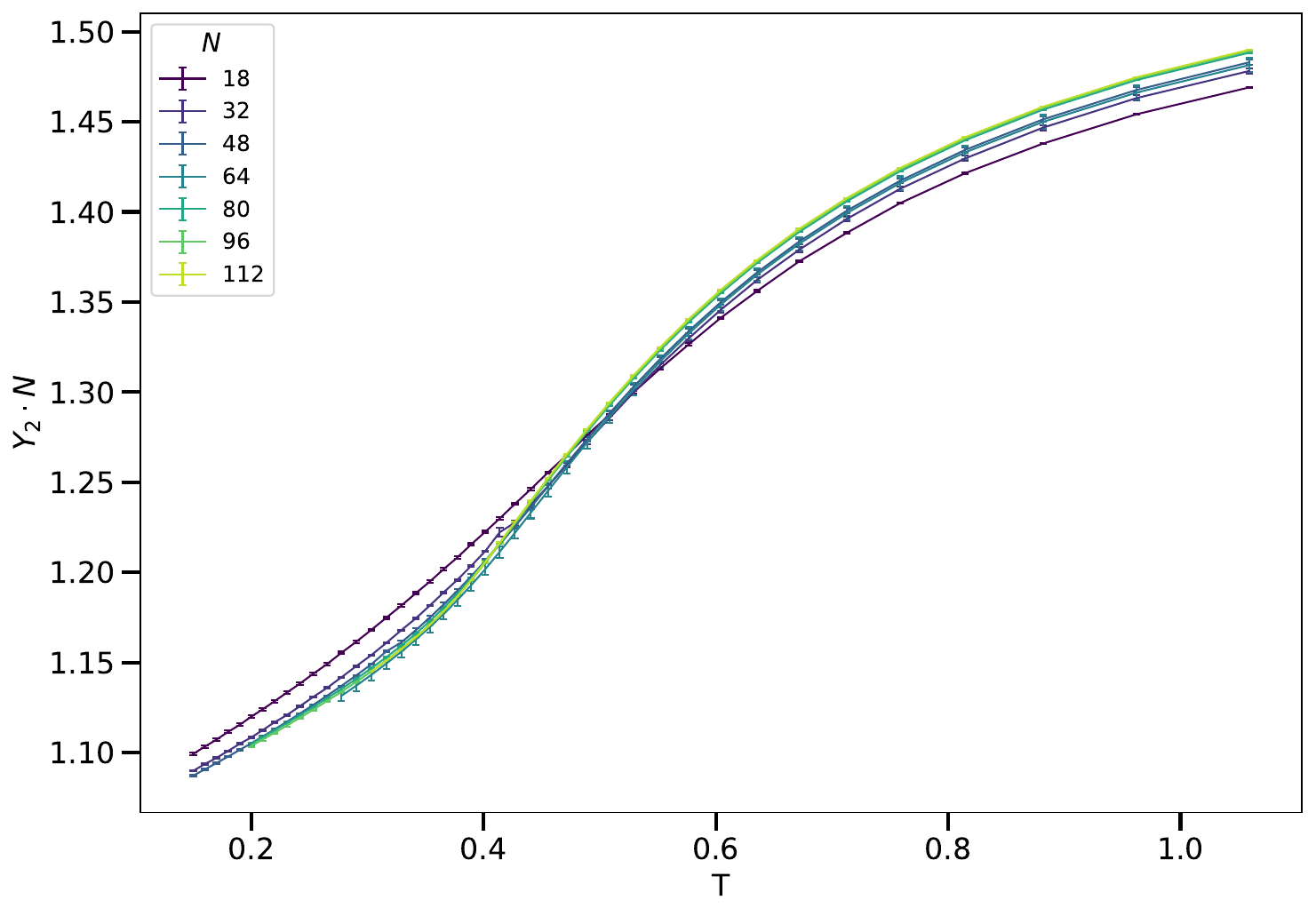}
    \caption{$Y_2(T)$ for the simulated sizes of the smoothed cubic random laser on the complete ML interaction graph.}
    \label{fig:Y2}
\end{figure}

\begin{figure}[t!]
    \centering
    \includegraphics[width=0.99\linewidth]{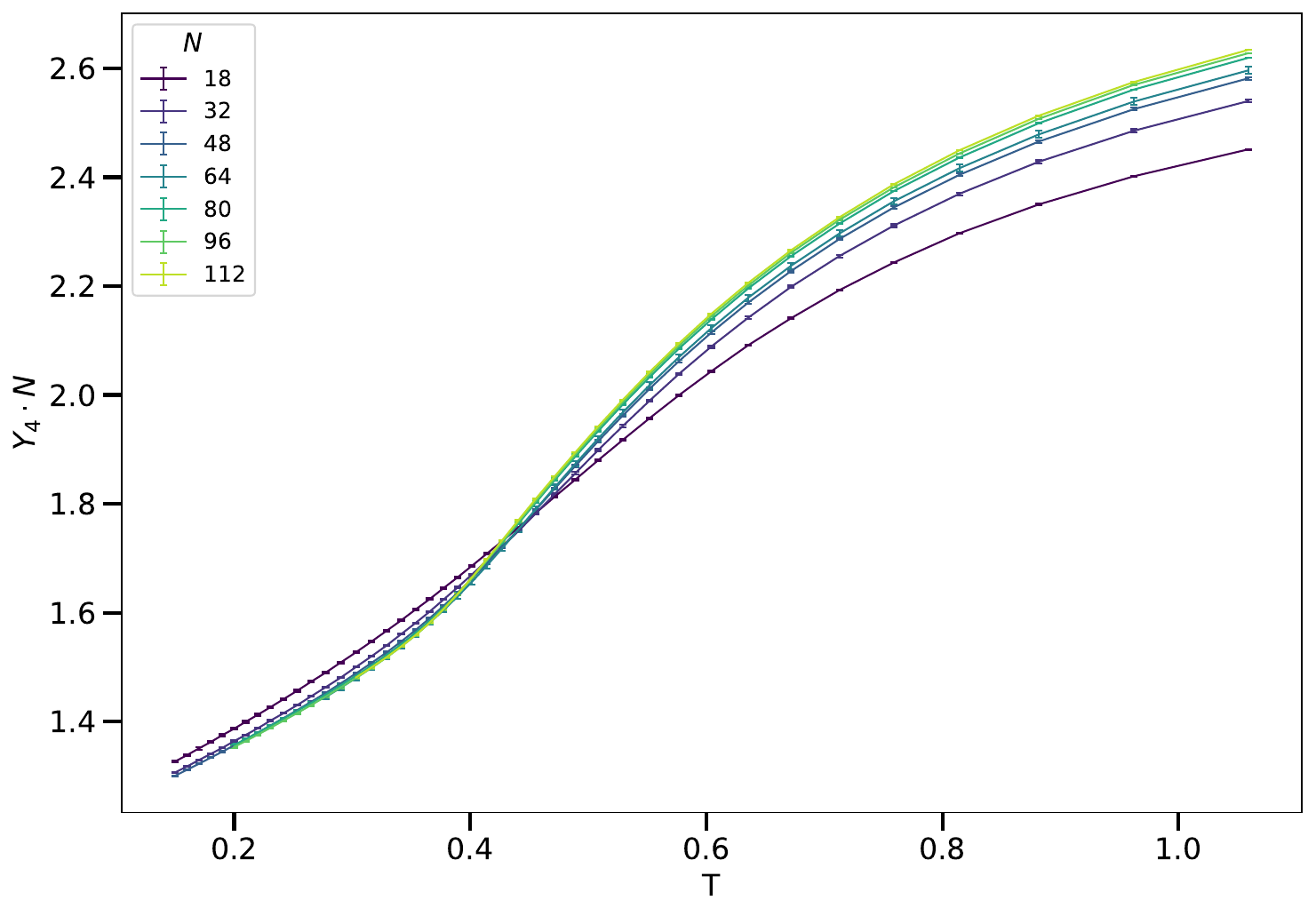}
    \caption{$Y_4(T)$ for the simulated sizes of the smoothed cubic random laser on the complete ML interaction graph.}
    \label{fig:Y4}
\end{figure}

\begin{figure}[t!]
    \centering
\includegraphics[width=0.49\linewidth]{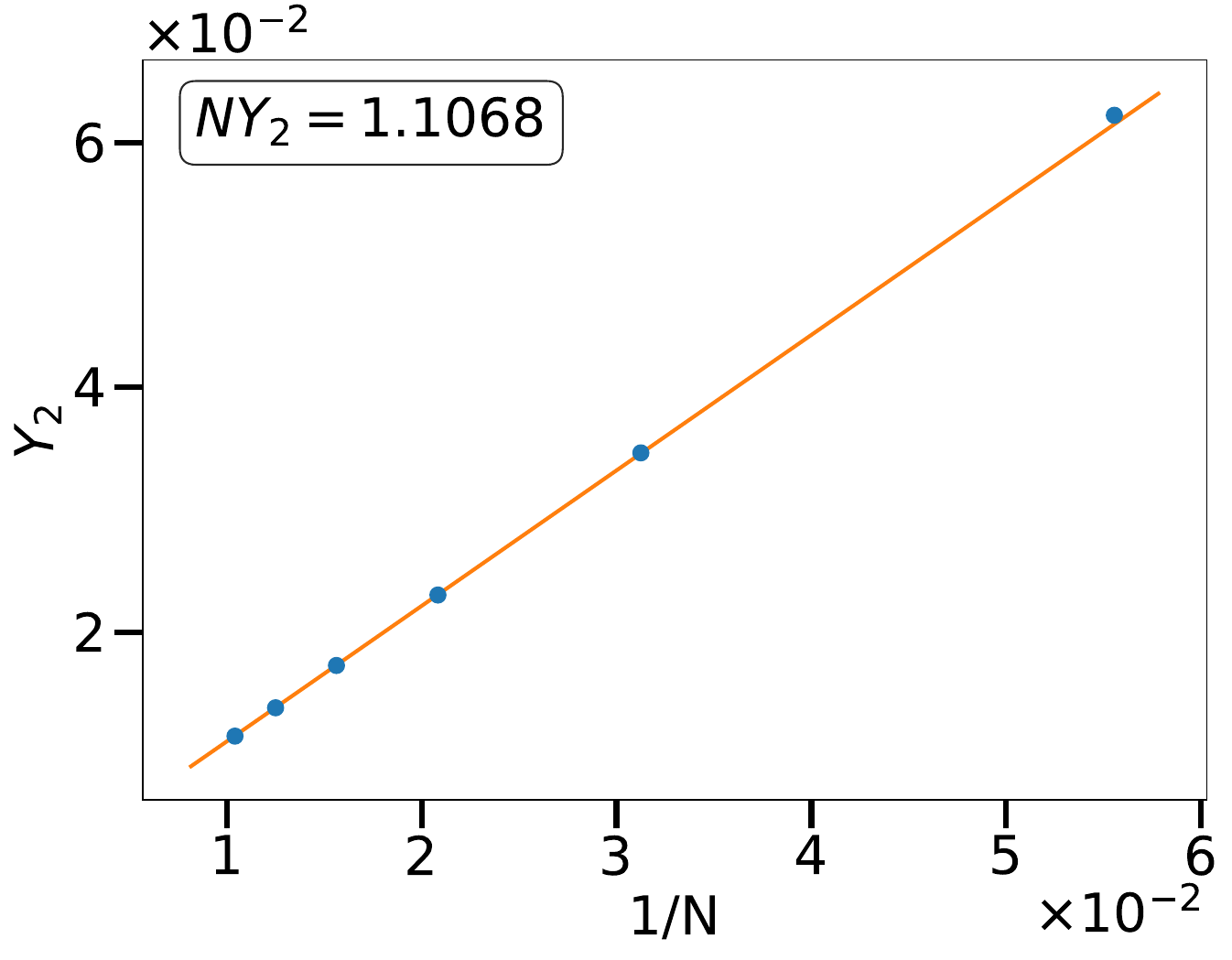}
    \includegraphics[width=0.49\linewidth]{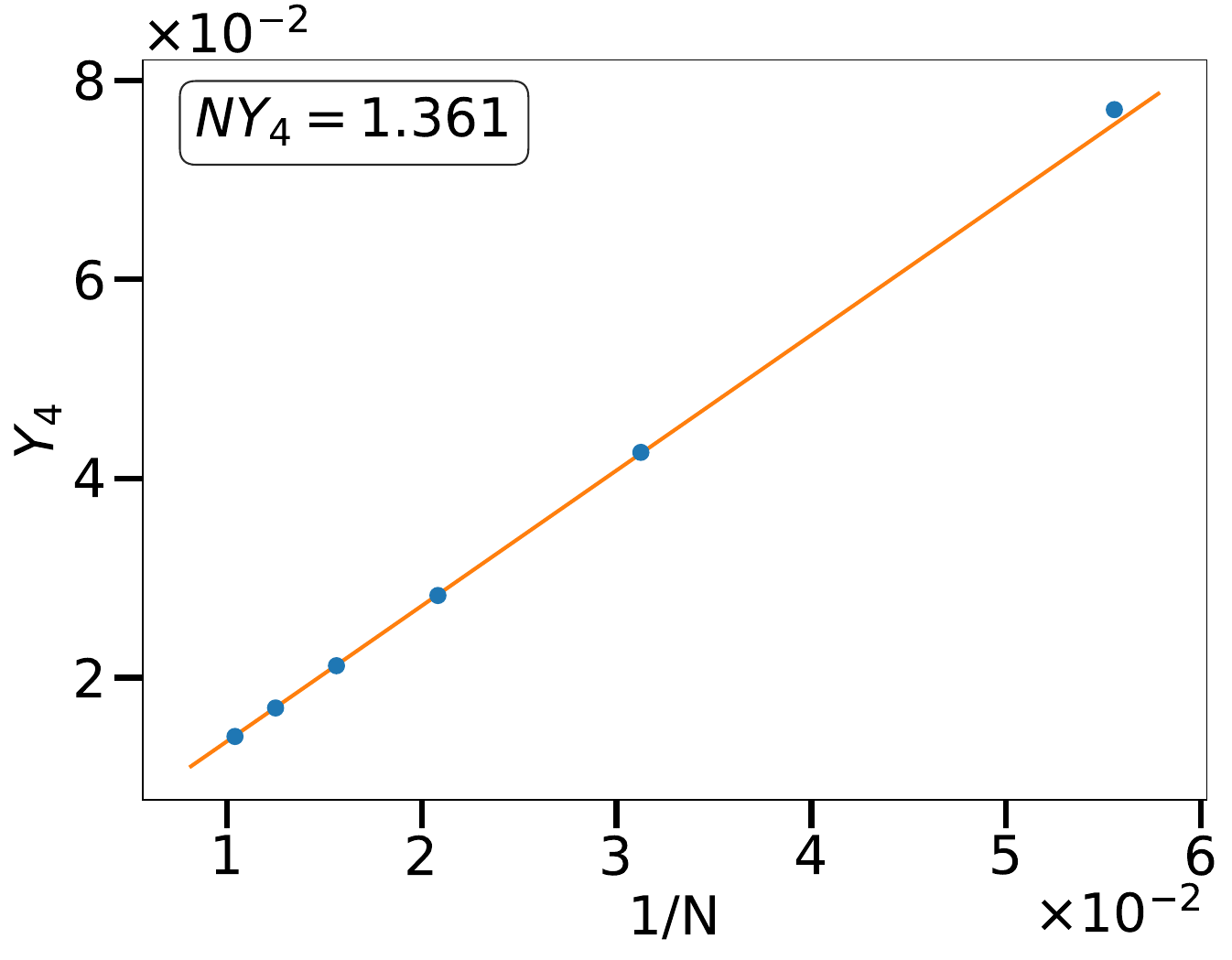}
    \caption{Scaling of $Y_n$ with $1/N$. Total intensity is spread over all modes:  $N Y_{2,4}$ is a constant for all $N$.  }
    \label{fig:Yscaling}
\end{figure}

\begin{itemize}
    \item If the total power pumped into the system is spread over all modes,  $|a_k|\sim 1\:\forall k$ and $\sum_k |a_k|^{r} \sim N $ for any exponent $r$. Therefore, it will be $Y_n \sim 1/N$ and a non-condensed system will be characterized by a constant $N\,Y_n$.
    
    \item If the system's power is condensed on a few modes,  there is only a finite number of spins $a_{k\in {\rm cond}}$ different from zero (possibly four, corresponding to a single interaction quadruplet). Because of the smoothed cubic constraint Eq. (\ref{SmoothedCubic}) they will scale like $|a_k| \sim N^{1/4}$. This implies that 
    $\sum_k |a_k|^r \sim |a_{k\in {\rm cond}}|^r \sim  N^{r/4}$ and $Y_n \sim O(1)$.
\end{itemize}
The reason why we implemented the analysis of both IPR's is that they are both needed to identify a possible pseudo-condensation in the smoothed cubic case.
However, this is not the case on the complete ML graph as we are now reporting.
From the FSS behavior of the IPR's we thus easily discriminate between  ($Y_n = {\rm const}$)  or spreading ($Y_n \sim 1/N \to 0)$.
\begin{eqnarray}
   (I)\ \,  & \mbox{ Intensity condensation: }   & Y_{2,4} \underset{N\to\infty }{\longrightarrow}  \mbox{const}
    \nonumber
    \\
 (II)  & \mbox{Intensity spreading:} &  Y_{2,4} \sim \frac{1}{N} \underset{N\to\infty}{\longrightarrow} 0
    \nonumber
\end{eqnarray}
In figures \ref{fig:Y2} and \ref{fig:Y4} we display the temperature behaviors of $NY_2(T)$ and $NY_4(T)$ for various sizes $N$. Both at high and low $T$ there is a scaling of the curves with the sizes, though it appears to converge quite rapidly, at least for the simulated sizes. To rule out the condensation, in Fig. \ref{fig:Yscaling} we plot the behavior of both $Y_2$ (left) and $Y_4$ (right) at temperature $T=0.2$. Both curves are  straight lines proportional to $1/N$. Therefore the intensity is spread among all modes. 

\begin{figure}
    \centering
\includegraphics[width=0.79\linewidth]{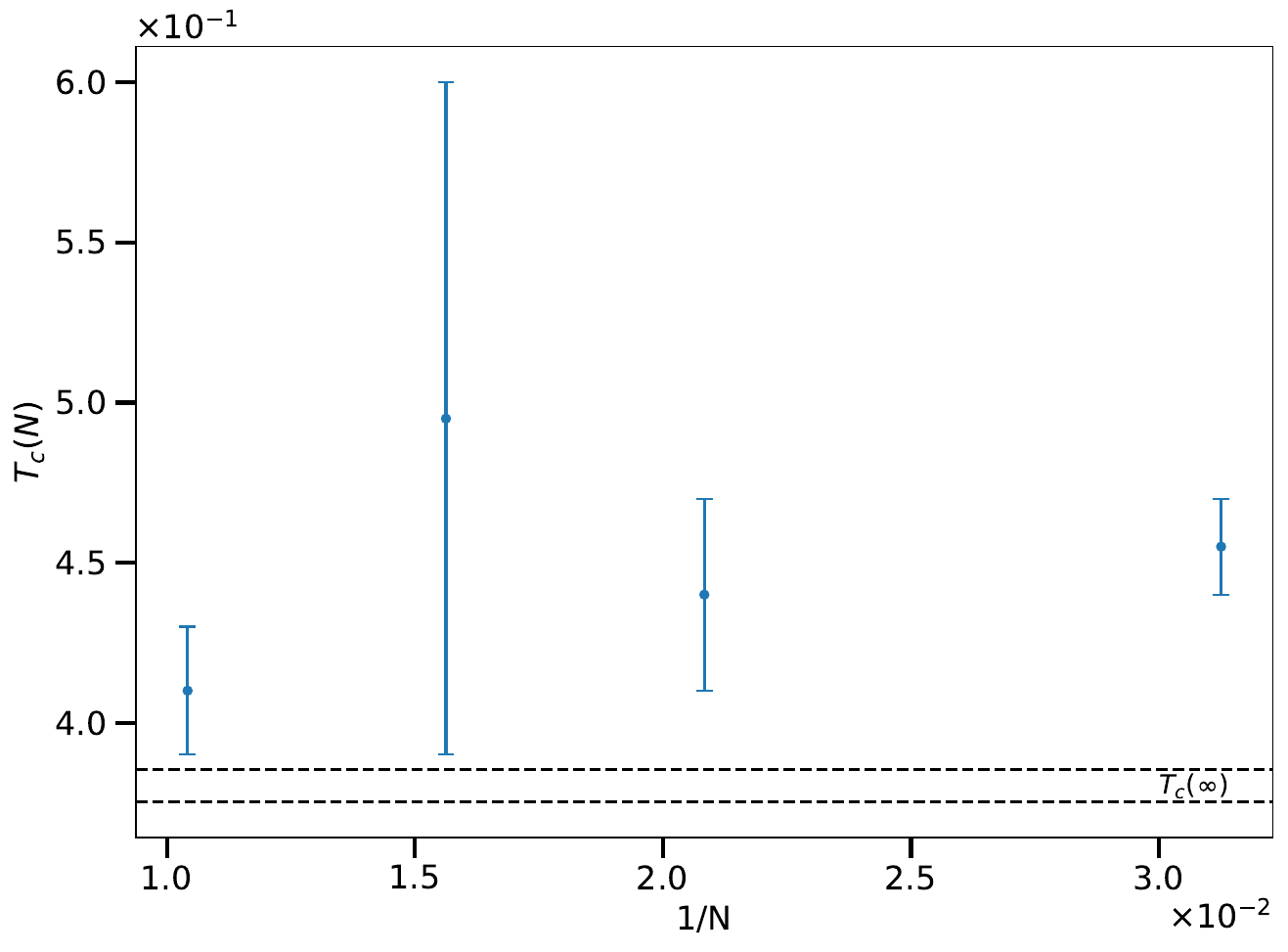}
    \caption{Crossing points of $NY_4(T)$ at nearby sizes. The dashed lines represent the interval of the FSS estimate of the critical temperature by means of specific heat data. }
    \label{fig:YTc}
\end{figure}

Incidentally, in Figs. \ref{fig:Y2}, \ref{fig:Y4} we can observe a tiny temprature region of scale invariance.
Also the IPR's, as the kurtosys of the $P(q)$ are scale invariant at criticality:

\begin{eqnarray}
\label{eq:yn_scaling_lattice}
    Y_{2,4}(T) = \hat f_{Y_{2,4}}\left(
    N^{\frac{1}{\nu_{\rm eff}}} t_N
    \right).
\end{eqnarray}
A clean finite scaling analysis 
of the crossing points is, unfortunately, unfeasible with the current data because the curves are too nearby with respect to the statistical errors obtained with the simulated samples. In Fig. \ref{fig:YTc} we display the crossing points of the $NY_4(T)$ curves of Fig. \ref{fig:Y4}. Though not completely  alien to the critical temperature estimate the present statistics, and size, is not enough to provide a reliable  $T_c$ estimate to compare. 

\section{Conclusions and outlook}
\label{sec:conc}
In this work, we have performed a detailed numerical investigation of a statistical mechanics model for random lasers, moving beyond the  spherical constraint to a more realistic smoothed cubic constraint, cf. Eq. (\ref{SmoothedCubic}). This model, defined on a dense, ML interaction network generated by the Frequency Matching Condition, Eq. (\ref{FMC}) captures the essential features of a $4$-body glassy random laser with a random frequency spectrum. 

Our GPU-accelerated Monte Carlo simulations, employing the Parallel Tempering algorithm, have allowed us to thermalize the system deep into the low-temperature, glassy phase for systems of up to $N=112$ modes. 

The  main results of our study are as follows. The system undergoes a phase transition from a high-temperature (low pumping) paramagnetic phase to a low-temperature (high pumping) glassy phase representing the random lasing regime. 
Finite-size scaling analysis of the specific heat peak allowed us to estimate the critical temperature in the thermodynamic limit, cf. Eq. (\ref{eq:Tc}).
The transition critical exponents estimates, see Eqs. (\ref{eq:alpha}) are compatible with those of the Random Energy Model (REM), suggesting that the universality class of this glassy random laser is mean-field. 

The analysis of the Parisi overlap distribution $P(q)$, furthermore, provides clear evidence of a multistate non-trival organization at low temperatures. As the temperature decreases below $T_c$, the $P(q)$ evolves from a single Gaussian peak to a distribution with developing side structures, signaling the emergence  of replica symmetry breaking. The scale-invariant behavior of the $P(q)$ kurtosis at the transition further corroborates the critical point identified from the specific heat.

A key technical and physical achievement of the smoothed cubic constraint is the effective suppression of intensity condensation. The analysis of the Inverse Participation Ratios (IPRs) $Y_2$  and $Y_4$ demonstrates that the total intensity is spread almost homogeneously over an extensive number of modes, in contrast to the pseudo-condensation phenomena observed in the spherical model on the same interaction network. This confirms the theoretical conjecture reported in Sec. \ref{sec:nocond}  that the constraint pushes the critical connectivity exponent for lack of power condensation $A_c$ to a lower value, cf. Eqs. (\ref{eq:con_scaling},\ref{Eq:Ac}), stabilizing a phase with distributed intensity.

In prospective, the property of the smoothed cubic constraint to prevent intensity condensation, even for sparsely connected networks, is currently being leveraged in work-in-progress studies that simulate models with finite connectivity per mode. This approach reduces computational demands and is enabling the investigation of larger systems, where the roles of network topology and power condensation can be further elucidated. Although the glass transition has been identified, a fuller resolution of the low-temperature phase -- particularly, the structure of the Parisi overlap distribution P(q) --remains an objective that might be more easily fulfilled working on a sparse network, acquiring more disordered samples statistics and longer simulation times. 

Finally, for a fully realistic comprehensive model, future work should reintroduce the linear interaction term $\mathcal H_2$ in Eq. (\ref{Hamilt2}), which accounts for net gain, losses, and the openness of the system. This would allow for a more direct comparison with the lasing threshold and the full emission characteristics of experimental random lasers.

In summary, by implementing a more physically motivated global constraint and leveraging advanced numerical methods, this work strengthens the conceptual bridge between random lasers and the statistical mechanics of disordered systems. It confirms the glassy character of random lasers and establishes an adaptable computational  framework for exploring their rich phenomenology further.

\section{Acknowledgements}
We thank  Maria Chiara Angelini, Giacomo Trinca Cintioli and Jacopo Niedda 
  for useful discussions. We thank Massimo Bernaschi for helping us with hardware support when we most needed it.
We acknowledge  funding from the Italian Ministry of University and Research, call PRIN 2022, project “Complexity, disorder and fluctuations”, grant code 2022LMHTET.  This study was conducted using the HPC infrastructure DARIAH of the National Research Council of Italy, by the CNR-NANOTEC in Lecce,
funded by the MUR PON “Ricerca e Innovazione 2014-2020” program, project code PIR01-00022.

\appendix
\section{Choice of temperatures}
\label{app:deltaT}
We chose the temperatures for the copies in the parallel tempering in order to have similar exchange rates between nearby copies. 
Temperatures between $0.2$ and $1$, around the candidate critical region, were chosen with the following criterium 
\begin{itemize}
    \item We simulated a system with size $N = 128$ and $2^{24}$ Monte Carlo steps, at fixed $\Delta T$ and with $125$ temperature replicas, and computed its specific heat $C_V(T)$. 
    
    \item We fixed three parameters $T_{\rm min}=0.2,\;T_{\rm max}=1,\;C=0.6$.
    
    \item We set $T_0 = T_{\rm min}$.
    
    \item We computed the temperature in the interval $[T_{\rm min},T_{\rm max}]$ recursively 
    \begin{equation}
       T_i = \frac{T_{i-1}}{1-\frac{C}{\sqrt{C_V(T_i)}}},
       \label{eq:PT_Ti}
    \end{equation}
       until $T_i < T_{\rm max}$.
\end{itemize}
Eq. (\ref{eq:PT_Ti}) is meant to satisfy the relation:
\begin{equation*}
    \Delta T \propto \frac{T}{\sqrt{C_V(T)}}
\end{equation*}
which should guarantee a (relatively) fixed acceptance rate of parallel temperature swaps at a value which depends on parameter $C$, see Refs. \cite{Marinari97,Kone05}. In our case, $C$ was chosen heuristically in order to facilitate thermalisation.
For sizes $N=80,96$ we used exactly this set of temperatures.
For sizes $N\in\{18, 32, 48, 64\}$ we simply added $5$ more equispaced temperatures $T\in\{1.5, 1.6, 1.7, 1.8, 1.9\}$, and for size $N=112$ we cut temperatures lower than $T=0.3$. In this way all sizes share the same temperatures in the common ranges and the acceptance rate is always almost homogeneous.

\section{Relationship between condensation and  interaction network connectivity in a multi-spin model with globally constrained  continuous spins}
\label{app:condensation}
Let us consider a generic $p$-spin interacting model with $N$ continuous spins $\sigma$ and a global constraint 
\begin{equation}
    \sum_{i=1}^N \sigma_i^\rho =N,
    \label{eq:SpherCon}
\end{equation} whose Hamiltonian is
\begin{equation}
    \mathcal H[\sigma] = -\sum_{k_1\ldots k_p}^{\# N^A} \sigma_{k_1}\ldots \sigma_{k_p} \ J_{k_1\ldots k_p}, 
    \label{eq:Hp}
\end{equation}
where $N^A$ on top of the sum, with $A\in [1,p]$,
 denotes the scaling with the size of the number of $p$-uples contributing to the energy and the spin indices  $k_{i}$ run from $1$ to $N$. 
 If $A=p$ we have a fully connected interaction graph, i.e.,~each single spin contributes in all other, $\mathcal O(N^{p-1})$, $p$-uples. 
 On the other extreme, if $A=1$ the graph is {\it sparse}, i.e.~each spin only interacts in a finite  number of $p$-uples, not growing with the size of the system. 
 All dilutions in between sparse and fully connected will be now considered. 
 Without any lack in generality, for simplicity we will take the variables as real valued here, rather than complex, yet  keeping the word {\it intensity} for the spin magnitude $|\sigma|$. 
 
 If intensity condensation occurs, that is, if only a few modes take the overall intensity, equal to $N$ according to  Eq.~(\ref{eq:SpherCon}), whereas all the other are zero,  what will be the energy contribution of a localized spin configuration?
First of all let us notice that in order to have a non-zero contribution the intensity of at least  $p$ coupled spins must localize. If we represent by $\square$ such a localizing $p$-uple the  intensity localized configuration is
\begin{equation}
    \{\sigma_{\rm loc}\}: \qquad \sigma \in \square \propto N^{1/\rho} \quad , \quad  \sigma \cancel{\in} \ \square =0.
    \label{eq:localization}
    \end{equation}

If the interaction couplings $J_{k_1\ldots k_p}$ are quenched disordered, independently distributed and with zero mean ${\overline{J_{k_1\ldots k_p}}}=0$, the typical ground state of the Hamiltonian (\ref{eq:Hp}) is extensive -- $E=\mathcal H [\sigma_{\rm gs}]=\mathcal O(N)$ --  if the variance of the distribution of the couplings scales like
\begin{equation}
    \label{eq:JNdisscale}
{\overline{J^2_{k_1\ldots k_p}}}\propto \frac{1}{N^{A-1}}.
\end{equation}
If the total intensity of the system condenses into a single interacting $p$-uple, as in (\ref{eq:localization}), Eqs. (\ref{eq:Hp}), (\ref{eq:JNdisscale}) imply that the energy scales with the size like
$$E_{\rm loc}=\mathcal H[\sigma_{\rm loc}] = 
\mathcal O\left(\frac{N^{p/\rho}}{N^{(A-1)/2}}\right).$$

To figure out whether intensity condensation might occur and dominate, one eventually has to compare the scaling behaviors of the energies of an intensity distributed and a condensed configuration.
Comparing  the distributed intensity contribution to the total energy $E=\mathcal O(N)$ and the condensed contribution $E_{\rm loc}$
we find the following three regimes as the exponent $A$ varies:

\begin{enumerate}
    \item $A> 2{p/\rho}-1$.  Energy contributions of possible condensed modes are subextensive. The distributed regime is dominant. The fully connected interaction graph case ($A=p$) lies in this regime if $\rho\geq 2$ (pedantically speaking $\rho>2-1/p$).
    
    \item $A= 2{p/\rho}-1$. Both kinds of spin configurations yield an $\mathcal O(N)$ contribution to the energy. 
        In this case one might conjecture the occurrence of a pseudo-condensed phase in which a few spins take an  extensive amount of the overall intensity, but leaving a finite fraction of it to be distributed among all other spins. This is the case for the $p=4$ spherical ($\rho=2$) case of the ML  random laser simulated in Ref. \cite{Niedda23b}. An analysis case by case has to be performed, though, because no argument can {\it a priori} exclude the occurrence of complete condensation or the total lack of it. 
    
    \item $A<2{p/\rho}-1$. Intensity condensation provides superextensive contributions to the energy and when the thermal noise is low enough condensation takes over. The equivalence between microcanonical and canonical ensemble breaks down.
    
\end{enumerate}

\bibliography{Lucabib}
\end{document}